\documentclass[letterpaper,12pt]{extarticle}

\usepackage{import}
\usepackage{local}
\usepackage[left=.75in,top=.75in,bottom=.75in,right=.75in]{geometry}

\usepackage[utf8]{inputenc}
\usepackage[T1]{fontenc}

\usepackage{float}
\usepackage{physics}
\usepackage{graphicx}% Include figure files
\usepackage{dcolumn}% Align table columns on decimal point 
\usepackage{bm}% bold math
\usepackage{mathrsfs}
\usepackage{amssymb}
\usepackage{amsmath}
\usepackage[normalem]{ulem}
\usepackage{wrapfig}
\usepackage[compress,square,sort,comma,numbers]{natbib}
\usepackage{csquotes}

\usepackage{caption}

\usepackage{hyperref}% add hypertext capabilities
\usepackage{bookmark}

\hypersetup{%
  breaklinks=true,
  colorlinks=true,
  linkcolor=black,
  filecolor=black,
  urlcolor=black,
  citecolor=black,
  pdfborder=0 0 0,
}

\usepackage{wrapfig}
\usepackage{tcolorbox}
\usepackage{framed}

\DeclareMathOperator{\sgn}{sgn}

\usepackage{xcolor}

\usepackage[normalem]{ulem}

\newtcolorbox{cellbox}[3][]{
width    = \textwidth,
colframe = #2!50,
colback = white,
coltitle = #2!20!black,
title    = { \large \textbf{#3}},
#1,
}

\begin{document}

%\title{Learning how gene patterns and closed-loop protein dynamics govern morphogenesis}
%\title{Learning how BMP controls the closed-loop dynamics of morphogenesis across species}
%\title{\LARGE Learning how BMP governs the dynamics of neuroectoderm morphogenesis across species}
\title{\LARGE Learning a conserved mechanism for early neuroectoderm morphogenesis}
%\title{How BMP plays a conserved role in the dynamics of morphogenesis}

\author{\textbf{Matthew Lefebvre\textcolor{Accent}{\textsuperscript{1,$\ast$}}, 
Jonathan Colen\textcolor{Accent}{\textsuperscript{2,3,$\ast$}}, 
Nikolas Claussen\textcolor{Accent}{\textsuperscript{1,$\ast$}},
Fridtjof Brauns\textcolor{Accent}{\textsuperscript{4}},
Marion Raich\textcolor{Accent}{\textsuperscript{5}},
Noah Mitchell\textcolor{Accent}{\textsuperscript{4}},
Michel Fruchart\textcolor{Accent}{\textsuperscript{2,3,6}},
Vincenzo Vitelli\textcolor{Accent}{\textsuperscript{2,3,7,8$\dagger$}},
Sebastian J Streichan\textcolor{Accent}{\textsuperscript{4,9,$\dagger$}}
} \\
\begin{small}
    \textcolor{Accent}{\textsuperscript{1}} Department of Physics, University of California, Santa Barbara, 93106, U.S.A. \\
    \textcolor{Accent}{\textsuperscript{2}} Department of Physics, University of Chicago, Chicago, Illinois, 60637, U.S.A. \\
    \textcolor{Accent}{\textsuperscript{3}} James Franck Institute, University of Chicago, Chicago, Illinois, 60637, U.S.A. \\
    \textcolor{Accent}{\textsuperscript{4}} Kavli Institute for Theoretical Physics, University of California, Santa Barbara, 93106, U.S.A. \\
    \textcolor{Accent}{\textsuperscript{5}} Center for Protein Assemblies and Lehrstuhl für Biophysik (E27), Physics Department, Technische Universität München, Garching, Germany \\
    \textcolor{Accent}{\textsuperscript{6}} Gulliver, ESPCI Paris, Université PSL, CNRS, 75005 Paris, France \\
    \textcolor{Accent}{\textsuperscript{7}}Kadanoff Center for Theoretical Physics, University of Chicago, Chicago, IL 60637, U.S.A. \\
    \textcolor{Accent}{\textsuperscript{8}}Institute for Biophysical Dynamics, University of Chicago, Chicago, IL 60637, U.S.A. \\
    \textcolor{Accent}{\textsuperscript{9}} Biomolecular Science and Engineering, University of California Santa Barbara, 93106, U.S.A \\
    \textcolor{Accent}{\textsuperscript{$\ast$}} These authors contributed equally \\
    \textcolor{Accent}{\textsuperscript{$\dagger$}} Correspondence: 
    \textcolor{Accent}{vitelli@uchicago.edu, streicha@ucsb.edu} \\
\end{small}
}

\maketitle

%%% NOMENCLATURE CONVENTIONS
% PROTEINS: myosin, E-cadherin, DPP.
% GENES: For cases where we discuss both the protein and the gene that codes for it, the protein is capitalized and the gene is not (Even-Skipped = protein, even-skipped = gene).
% ALLELES: \emph{gene^{allele}}, gene capitalized only if the allele is dominant
% ABBREVIATIONS: ML (check), NN (check), DV (check), AP (check), PRG (check), VF (check), GBE (check).  - make sure all of them are defined

\begin{abstract}

Morphogenesis is the process whereby the body of an organism develops its target shape. The morphogen BMP is known to play a conserved role across bilaterian organisms in determining the dorsoventral (DV) axis. Yet, how BMP governs the spatio-temporal dynamics of cytoskeletal proteins driving morphogenetic flow remains an open question. Here, we use machine learning to mine a morphodynamic atlas of Drosophila development, and construct a mathematical model capable of predicting the coupled dynamics of myosin, E-cadherin, and morphogenetic flow. Mutant analysis shows that BMP sets the initial condition of this dynamical system according to the following signaling cascade: BMP establishes DV pair-rule-gene patterns that set-up an E-cadherin gradient which in turn creates a myosin gradient in the opposite direction through mechanochemical feedbacks. Using neural tube organoids, we argue that BMP, and the signaling cascade it triggers, prime the conserved dynamics of neuroectoderm morphogenesis from fly to humans.
%argue for a conserved role of BMP in priming homeobox genes and cytoskeletal dynamics to drive morphogenesis across animal species, from flies to human.

\end{abstract}

\newpage

\section{Introduction}

Morphogenesis is the process by which the shape of an organism emerges from the coordinated behavior of groups of cells.
Turing's pioneering work traced morphogenesis to the presence of \enquote{chemical substances called morphogens, reacting together and diffusing through a tissue}~\cite{TuringAlan1952}. 
Molecular biology has since identified these morphogens as molecules whose concentration modulates the expressions of various genes~\cite{Gilmour2017}. 
%This in turn determines a cell's fate -- namely, its future identity or that of its daughter cells.
During early development, morphogen concentrations set up a spatial coordinate system and body axes on which the embryo organizes future tissues and organs~\cite{Sheng2021}.
However, it is force-generating proteins -- not morphogens --  that produce the active mechanics which moves each group of cells into the right place~\cite{Marchetti2013,Prost2015,Saarloos2023, deneke_self-organized_2019, saadaouiTensileRingDrives2020}.
%Active mechanics describes how biology harnesses such physical processes to create shape
%A central challenge in developmental biology is to unveil general mechanisms by which shapes are generated.
This raises the questions: \textit{How do morphogens control the active mechanics that execute morphogenesis? And are these mechanisms conserved across species?} 
%This raises the question: \textit{How do morphogens control the active mechanics that execute morphogenesis?}

%A central challenge in developmental biology is to unveil general mechanisms by which shapes are generated.
%morphogens control the active mechanics of shape-creating proteins. 
We focus on the paradigmatic example of neuroectoderm morphogenesis, which is the first step in forming the central nervous system in species ranging from \emph{Drosophila melanogaster} to \emph{Homo sapiens}.
This process is regulated by bone morphogenetic protein (BMP) signaling, whose role as a morphogen is conserved across organisms with bilateral symmetry~\cite{Bier2011,Bier2015}. 
In addition, tissue elongation is driven by a second conserved mechanism known as convergent extension~\cite{Pare.Zallen2020}.
However, the exact process by which BMPs control tissue mechanics is unknown.

% 1. Cell positions vary from embryo to embryo, while a coarser perspective is repeatable, just like morphogenesis
The dynamics of these tissues is dauntingly complex. Nonetheless, a remarkable simplification occurs when focusing on large scale patterns. %, while morphogenesis is repeatable.
To obtain consistent and reproducible dynamics across different embryos, we coarse-grain data, i.e. we smooth it in space and align it in time~\cite{mitchellMorphodynamicAtlasDrosophila2022}. 
% 2. Recent advances have enabled a huge amount of coarse-grained data
Recent advances in light-sheet microscopy~\cite{krzicMultiviewLightsheetMicroscope2012} have enabled the collection of such coarse-grained data for tissue and protein dynamics across the entire embryo~\cite{mitchellMorphodynamicAtlasDrosophila2022}. 
% 3. Even taking prior insight, we're left with a difficult question: what do we need to predict mechanics?
At the cellular level, tissue deformations arise from the interaction of many biomolecules, as exemplified in Box 1 for \emph{Drosophila}. %including myosin, cell adhesion, and PRGs. 
At the coarse-grained level, a smaller number of these correlated quantities may be sufficient to predict the large-scale mechanical behavior.
% 4. To resolve this question in a model-free way, we turn to ML
In order to identify the minimal biochemical information that allows one to encode tissue flow, we use statistical inference tools such as machine-learning (ML)~\cite{lecunDeepLearning2015,mehtaHighbiasLowvarianceIntroduction2019,carleoMachineLearningPhysical2019}, as shown in Fig.\ref{fig:deepNN}e-f.
% 5. ML alone establishes correlation, but not causation. 
%By themselves, these methods do not distinguish correlation from causation.
These techniques alone are not sufficient to learn causal rules for tissue dynamics.
% 6. To understand morphogenesis, we would like causal rules that we can understand
% 7. By supplementing/supporting ML with insights from bio/physics, we can learn interpretable causal rules governing mechanics of convergent extension
%In order to learn causal rules governing convergent extension, 
To do so, we must supplement ML with insights from biology and physics~\cite{karniadakis_physics-informed_2021,murdoch2019pnas,Schmitt.etal2023}. 
% 8. What we do
Here, we develop a pipeline that integrates \emph{in toto} light-sheet microscopy, interpretable machine learning, and physical theories of active mechanics in order to develop and test a biological hypothesis: 
BMP shapes the dynamics of convergent extension %in both the human and the fly 
by initiating mechanochemical feedback loops involving force-generating proteins as well as genetic regulation. %PRGs.

This paper is organized as follows. 
We first analyze convergent extension in \emph{Drosophila} embryos.
Using our integrated pipeline, we identify predictive biophysical quantities --- tissue flow, myosin, and adherens junctional proteins --- and causal rules which describe their coupled dynamics during a key tissue deformation that occurs during Drosophila development, known as germ-band extension (GBE)~\cite{irvineCellIntercalationDrosophila1994}.
We then test the validity and generalizability of these rules using mutant analysis.
%and reveal how BMP controls these active mechanics by establishing an initial DV gradient in cell adhesion. 
In a nutshell, BMP establishes a global cell adhesion pattern which modulates mechanochemical feedbacks to produce a myosin contractility gradient in the opposite direction. 
Finally, we turn to experiments on human stem cell-based neural tube organoids and unveil how BMP, and the signaling cascade it triggers, prime the conserved dynamics of neuroectoderm morphogenesis from flies to humans.
\definecolor{boxbordercolor}{HTML}{C7BEBA}

%\begin{cellbox}{blue}{Box 1. Juxtaposition model of \emph{Drosophila} germ-band extension}
%    \footnotesize
%    \input{Boxes/BoxSebastian}
%\end{cellbox}

\section{Results}

\subsection{Neural networks forecast tissue dynamics from initial myosin distribution}

\begin{figure}
    \centering
    \includegraphics[width=\textwidth]{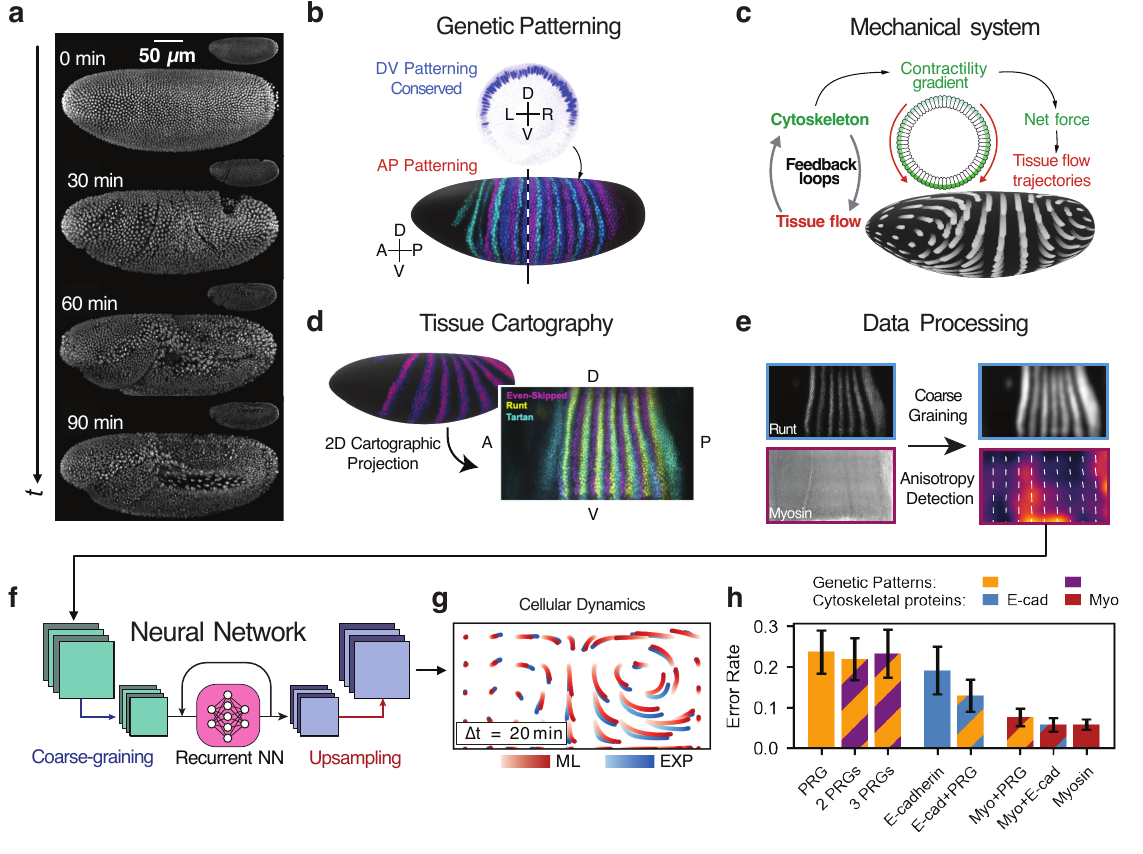}
    \caption{\textcolor{Highlight}{\textbf{Drosophila GBE as a system for understanding the interplay between genetic patterning and cytoskeletal force generation using machine learning.}}
    (\textbf{a}) Snapshots of nuclei from \emph{in toto} live recording of 90 minutes of \emph{Drosophila} gastrulation, comprising invagination of the ventral furrow and GBE.
    (\textbf{b}) %(Top, cartoon) 
    The conserved morphogen DPP (BMP) %(the \textit{Drosophila} homolog of BMP) 
    patterns cell fates along the DV axis, fates illustrated in blue. %(Bottom) 
     The AP axis is patterned by striped expression of PRGs, three of which are shown (Even-skipped: blue, Runt: cyan, Paired: magenta, virtual overlay created from 3 different embryos). 
    (\textbf{c}) A gradient of myosin recruitment along the DV axis generates forces that drive tissue flow.  Mechanical feedback loops respond to tissue deformation, sculpting local patterns of cytoskeletal recruitment.
    (\textbf{d}) Tissue cartography projects the curved 3D embryo surfaces onto a 2D plane. (\textbf{e}) Additional image processing includes spatial downsampling/smoothing and optional anisotropy detection via a radon transform algorithm. Anisotropy detection yields a tensor field, which can be thought of as a double-headed arrow whose direction aligns with cell junctions where a protein lies and whose magnitude reflects the local protein density. 
    (\textbf{f}) A NN forecasts tissue dynamics from the measured biological initial conditions. The network is a residual autoencoder which maps the inputs to a latent vector, predicts dynamics using a recurrent layer, and then translates the latent vector sequence into flow fields.
    (\textbf{g}) Trajectories of test points in flow fields measured in experiment (blue) and predicted by NN from an initial snapshot of myosin (red).
    (\textbf{h}) Forecasting performance of NNs trained on genetic and cytoskeletal patterns. NNs trained on myosin achieve the lowest error rates.
    }
    \label{fig:deepNN}
\end{figure}

We first aim to isolate the smallest set of proteins that allow us to forecast mechanical behavior from initial conditions.
To do so, we leverage a morphodynamic atlas of \emph{Drosophila} gastrulation which contains \textit{in toto} protein expression movies for various patterning genes and components of the cytoskeleton~\cite{mitchellMorphodynamicAtlasDrosophila2022} (Fig.~\ref{fig:deepNN}a-e).
We then train deep neural networks (NNs) to learn tissue dynamics (see SI for details) using different combinations of these movies. This agnostic approach allows us to identify predictive biochemical quantities without specifying a physical or biological model beforehand (Fig.~\ref{fig:deepNN}e-f). For our purpose, it is actually advantageous that neural networks are essentially input-output black boxes agnostic to the biological rules that we are yet to discover. 
We find that knowledge of the initial myosin distribution ($t=0$ on Fig.~\ref{fig:deepNN}a corresponding to ventral furrow formation) is sufficient for a NN to learn to forecast tissue flow in wild type (WT) embryos for 20 minutes and maintain excellent agreement with experiments through the onset of GBE (Fig.~\ref{fig:deepNN}g).
Previous work has evidenced an instantaneous relation between myosin and tissue flow during convergent extension~\cite{streichanGlobalMorphogeneticFlow2018}. Our results go further, suggesting that myosin contains enough information to quantitatively account for the dynamics of cellular flow and resulting trajectories (Fig.~\ref{fig:deepNN}g).
Finally, comparison of NNs trained on different sets of proteins indicates that the myosin distribution is necessary to forecast the flow from an initial condition: Fig.~\ref{fig:deepNN}h shows that any other protein in the atlas (or combination thereof) produces a higher prediction error.

\subsection{Coarse-grained dynamics of GBE are low-dimensional and reproducible across embryos}

Neural networks can perform forecasting, but are ultimately a black box from which it is difficult to infer interpretable causal rules. 
To gain further insight into the dynamics of myosin and flow during GBE, we perform a principal component analysis (PCA) of the coarse-grained fields and track the time evolution of the weight of each principal component over the entire duration of GBE (Fig.~\ref{fig:PCA}a).
In a nutshell, PCA decomposes the data into characteristic features known as \enquote{principal components} (PCs).
The weight of a PC describes how much this component contributes to the data.
As illustrated in Fig.~\ref{fig:PCA}b, we consider a fixed set of PCs and encode the time evolution of the data into time-dependent weights $a_i(t)$ of the PCs $PCi$.
We find that a small number of PCs can account for the dynamics of proteins and tissue flow during GBE (Fig.~\ref{fig:PCA}c).
Moreover, the dynamics in the low-dimensional spaces represented by the primary PCs -- the dominant patterns -- are simple and repeatable across embryos (Fig.~\ref{fig:PCA}a, fourth column).
The PCs give us insight into the dynamics: there is little myosin at early times, but the weight of the first PC increases as gastrulation proceeds, indicating development of a DV-patterned myosin field.
Similarly, there is little flow at early times, but the weight of the first PC increases during the onset of GBE where a characteristic flow pattern with four vortices develops. 
The simplicity and reproducibility across embryos revealed by PCA suggest that simple rules for GBE can be obtained.

\subsection{Active matter models of GBE require a patterned control field}

Having identified the most crucial features of the myosin pattern and tissue flow, we attempt to reproduce their dynamics using minimal mathematical models inspired by active matter theories that successfully describe cell and tissue mechanics~\cite{Prost2015,Marchetti2013,turlier_furrow_2014}.

%We rule this out because it would lead to randomly-positioned patterns, in contradiction with experimental observations where the myosin concentration is always highest in the ventral region.

\begin{figure}
    \centering
    \includegraphics[width=0.8\textwidth]{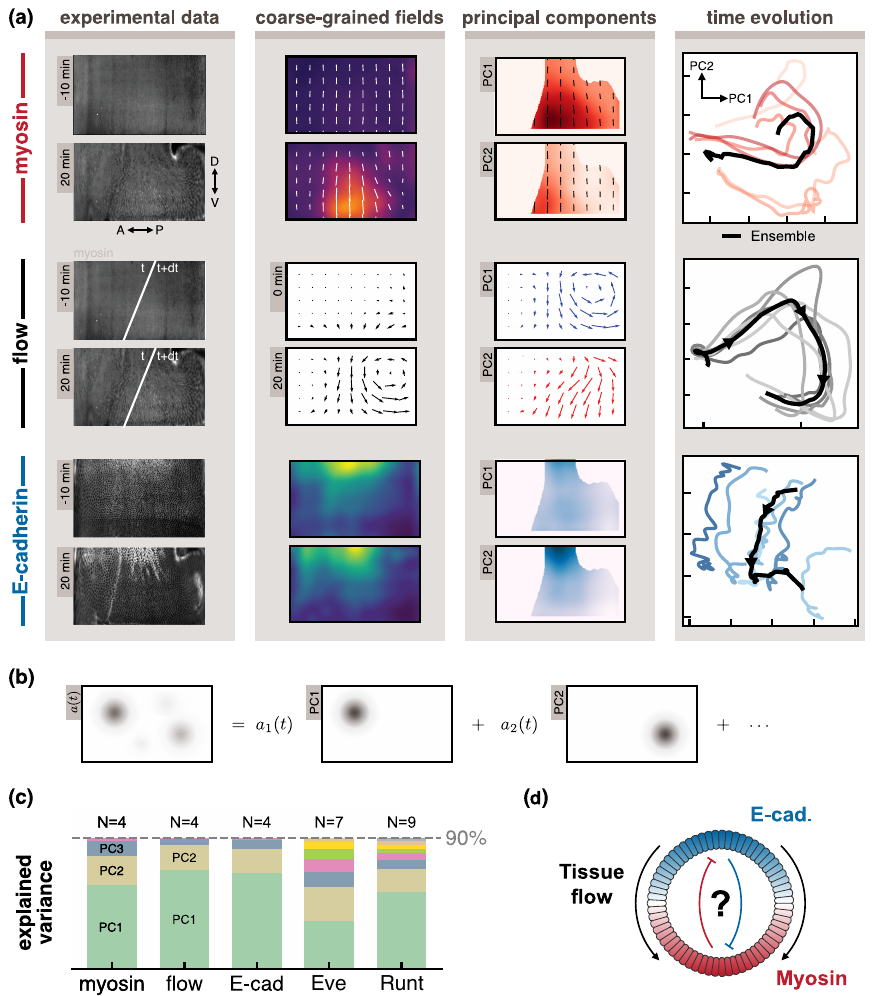}
    \caption{\textcolor{Highlight}{\textbf{Coarse-grained dynamics of GBE are low-dimensional and reproducible across embryos.}}
        (\textbf{a})
        (first column) Representative fluorescence images from experimental movies. Times are relative to initiation of the ventral furrow at $t=0$ min. Flow is computed via PIV on the fluorescence images.
        (second column) Myosin is coarse-grained using a radon transform algorithm into a tensor field that captures density and junctional orientation. Tissue flow is a vector field obtained via PIV. E-cadherin is coarse-grained into a scalar density field by smoothing the fluorescence image. 
        (third column) Primary PCs for myosin and E-cadherin are DV-graded in opposite directions. Here we mask out cephalic furrow and posterior midgut regions which become prominent at later times.
        Primary PCs for tissue flow describe the positions of vortices in the GBE flow field. 
        (fourth column) Projection of dynamics onto the space spanned by the primary PCs demonstrates smooth and reproducible behavior across embryos.
        (\textbf{b}) PCA decomposes a complex pattern into a number of characteristic features, or principal components. 
        (\textbf{c}) We quantify the complexity of a field by the number of principal components required to describe 90\% of dataset variation. Myosin, tissue flow, and E-cadherin require 4 PCs, while PRGs (Runt and Even-Skipped) require more. 
        (\textbf{d}) The relation between the PCA of myosin and E-cadherin suggests their dynamics may be coupled. 
        }
    \label{fig:PCA}
\end{figure}

It is challenging to find a dynamical system including only myosin and flow that reproduces long-lived dynamic states of development observed in experiments (Fig.~\ref{fig:PCA}a).% and Fig.~\ref{fig:MyosinOnly}-\ref{fig:MyoCad}).
The challenge mainly consists in obtaining a myosin gradient from initially nearly uniform myosin and velocity fields (at $t=0$, see Fig.~\ref{fig:PCA}a). 
As myosin is always higher on the ventral pole, we conclude that it must be modulated by some field that exhibits a gradient at early times (and exclude a pattern-forming instability). 
Such a ``control'' field, which may be established by upstream regulation, induces a myosin gradient along the DV axis, which in turn causes a GBE-like flow with four vortices. 
In addition, we find that a mechanical feedback loop that couples tissue deformation to myosin recruitment (see Box 1 and Refs.~\cite{Fernandez-Gonzalez2009,Yu2016,streichanGlobalMorphogeneticFlow2018,gustafsonPatternedMechanicalFeedback2022,lefebvre_geometric_2023, nishizawa_two-point_2023}) can extend the lifetime of this quasi-stationary state, which would otherwise be destroyed by nonlinear effects such as advection and junction rotation.% (see Fig.~\ref{fig:orientation_dynamics}-\ref{fig:mech_feedback}).

\begin{cellbox}{boxbordercolor}{Box 1. \emph{Drospohila} morphogenesis: a primer}
    \footnotesize
    \begin{wrapfigure}[24]{r}{0.7\textwidth}
    \includegraphics[width=0.68\textwidth]{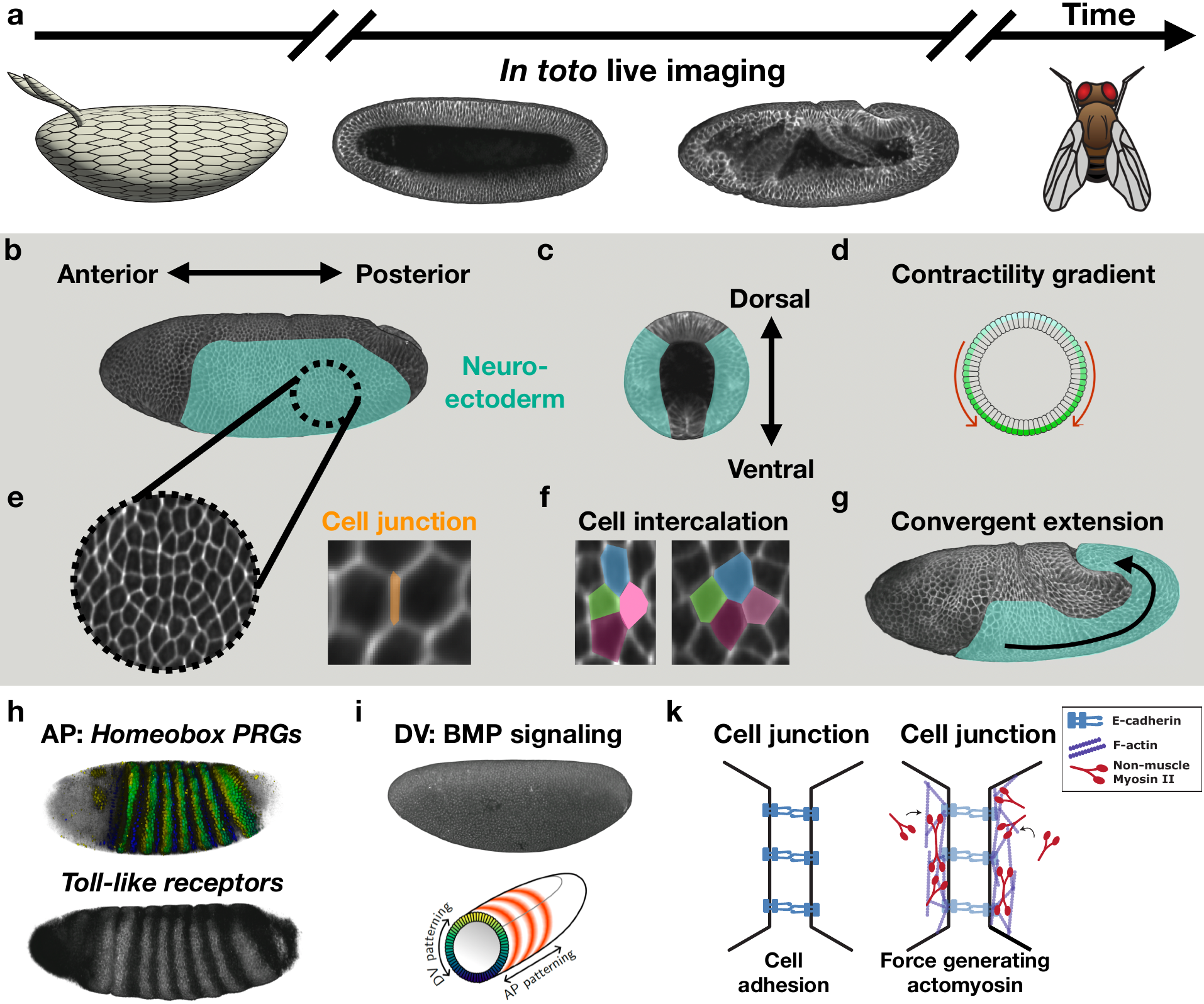}
\end{wrapfigure}
%To bridge this gap, we first focus on a well-studied model system: \emph{Drosophila}.
In \emph{Drosophila}, convergent extension drives a major morphogenetic movement known as germ-band extension (GBE)~\cite{irvineCellIntercalationDrosophila1994}, in which the neuroectoderm elongates along the anterior-posterior (AP) axis (\textit{right}, a-g).
This process arises through organized cell intercalations, in which cell interfaces oriented along the dorsal-ventral (DV) axis contract and are replaced by ones oriented along the anterior-posterior (AP) axis (\textit{right}, e-f).  
The mechanics at cell interfaces (\textit{right}, k) is driven by the interplay between force-generating proteins such as non-muscle myosin-II (myosin), a motor protein which causes junctions to contract, and the adherens junction complex which involves E-cadherin and F-actin to build the substrate for myosin. %resists the ensuing cell rearrangement. 
It has been proposed that transcription factors known as pair-rule-genes (PRGs) organize germ-band extension by regulating a family of receptors, known as Toll-like receptors~\cite{Zallen2004,Pare2014}, located at cell junctions (\textit{right}, h). 
% transmembrane receptors
PRGs are patterned in periodic stripes along the AP axis (\textit{right}, h), and could therefore provide a directionality to the process by modulating the recruitment of force-generating proteins depending on the orientation of their interfaces.
On the other hand, it has been shown that 
junctional myosin contraction
%myosin 
exhibits a gradient along the DV axis (\textit{right}, d), which is sufficient to predict the tissue flow during convergent extension~\cite{streichanGlobalMorphogeneticFlow2018}.
However, this global myosin gradient 
is stationary, and
cannot be attributed to PRG patterns
that rotate with the tissue~\cite{lefebvre_geometric_2023} 
but can be explained by mechanochemical feedback loops~\cite{gustafsonPatternedMechanicalFeedback2022,brauns_epithelial_2023}.
The outstanding challenge therefore becomes: how are the AP and DV axes (\textit{}{right}, h-i) integrated? Understanding the dynamics of myosin holds the key to address this question. 
%to quantitatively account for the time-course of the global myosin pattern.
\end{cellbox}

\subsection{E-cadherin acts as a patterned control field for GBE}

We now aim to identify a biochemical candidate for the control field that establishes the myosin recruitment gradient.  
As shown in figure \ref{fig:PCA}a, E-cadherin expression harbors some of the features expected of such a control field.
In particular, E-cadherin exhibits a DV gradient before the onset of GBE (Fig.~\ref{fig:PCA}a, bottom row), and could therefore serve as a control field for GBE.
%Biological reasons
The mechanics of tissue flow arises from the behavior of cell junctions, where myosin is recruited and generates forces (Box 1).
E-cadherin couples adjacent cells to each other at cell junctions, and is critical to establish the substrate to which the actin cytoskeleton is anchored. Moreover, E-cadherin has been proposed as a regulator of mechanotransduction in cells \cite{priya_active_2015, noordstra_cadherins_2023}. 
Motivated by the observation that other junctional proteins have similar expression profiles (Fig.~\ref{fig:bazooka}), % (Fig.~\ref{fig:bazooka}), 
we propose that E-cadherin can be used as a proxy to represent the overall state of cell junctions.
In addition, high levels of E-cadherin have been correlated with an inhibition of junctional recruitment of myosin both in tissues as well as individual cells ~\cite{Sumi.etal2018, truong_quang_extent_2021}.

% Information theoretical reasons
At the same time, we find that the number of PCs needed to represent the dynamics of E-cadherin is identical to that of myosin and tissue flow (Fig.~\ref{fig:PCA}c). 
The relation between the PCA of myosin and E-cadherin suggests their dynamics may be coupled (Fig.~\ref{fig:PCA}d). 
The primary PCs of E-cadherin are DV-graded in the opposite direction of myosin (Fig.~\ref{fig:PCA}a) and undergo similar dynamics. 
We note that in other contexts, morphogen gradients in opposite directions are known to shape stable gene expression patterns~\cite{Dubuis2013,Morishita2009,Zagorski2017,Krotov2014,Vetter2022,BenZvi2011,BenZvi2010}.

We could also have considered other upstream regulators such as PRGs (Box 1). 
Indeed, mutations to certain PRGs lead to abnormal myosin recruitment and tissue flow during GBE~\cite{irvineCellIntercalationDrosophila1994,Zallen2004,Fernandez-Gonzalez2009,lefebvre_geometric_2023}.
However, we find that PRG patterns contain much more complex information (Fig.~\ref{fig:PCA}c). This can be traced to the observation that PRGs flow with the tissue: their stripes continuously change their orientation, leading to an orientation discrepancy with myosin, whose orientation is nearly stationary~\cite{lefebvre_geometric_2023}. 
%...PRG patterns are continuously deformed by tissue flow, whereas myosin appears nearly stationary~\cite{lefebvre_geometric_2023}.
This motivates us to first focus on a post-translational picture.%, involving only elements that exist after RNA translation.

%As their descriptions require more PCs (Fig.~\ref{fig:PCA}b), it is unlikely that they are directly modulating the simpler (lower-dimensional) myosin dynamics. 
%that a direct signaling pathway exists between them providing 
%Our information-theoretic argument to decouple the dynamics of PRGs and force-generating proteins aligns with previous work finding that 

\subsection{Machine learning yields interpretable rules for neuroectoderm morphogenesis}

At this stage, our tentative picture is as follows:
E-cadherin acts as a control field to recruit myosin, which puts cells into motion.
We now aim to assemble these ingredients into a mathematical model that describes experimental data.
Using a ML technique known as  SINDy~\cite{bruntonDiscoveringGoverningEquations2016}, we learn interpretable biophysical equations for the coupled dynamics of myosin and E-cadherin (see Box 2).
These equations represent rules and feedback mechanisms which ultimately govern \emph{Drosophila} neuroectoderm morphogenesis. 
These rules involve several mechanisms
%--
%E-cadherin-modulated myosin detachment as well as strain-rate and tension recruitment (see Box 2 and SI).
that we traced back to existing experiments: myosin detachment~\cite{lefebvre_geometric_2023}, tension recruitment~\cite{lefebvre_geometric_2023,brauns_epithelial_2023}, and strain-rate feedback~\cite{gustafsonPatternedMechanicalFeedback2022,nishizawa_two-point_2023}.
Crucially, we find that the strength of each of these mechanisms is modulated by the local E-cadherin concentration. % For a cadherin pattern $c$, prefactors of $(1-c)$ are responsible for establishing a myosin DV gradient opposite that of cadherin.

\begin{figure}
    \centering
    \includegraphics[width=\textwidth]{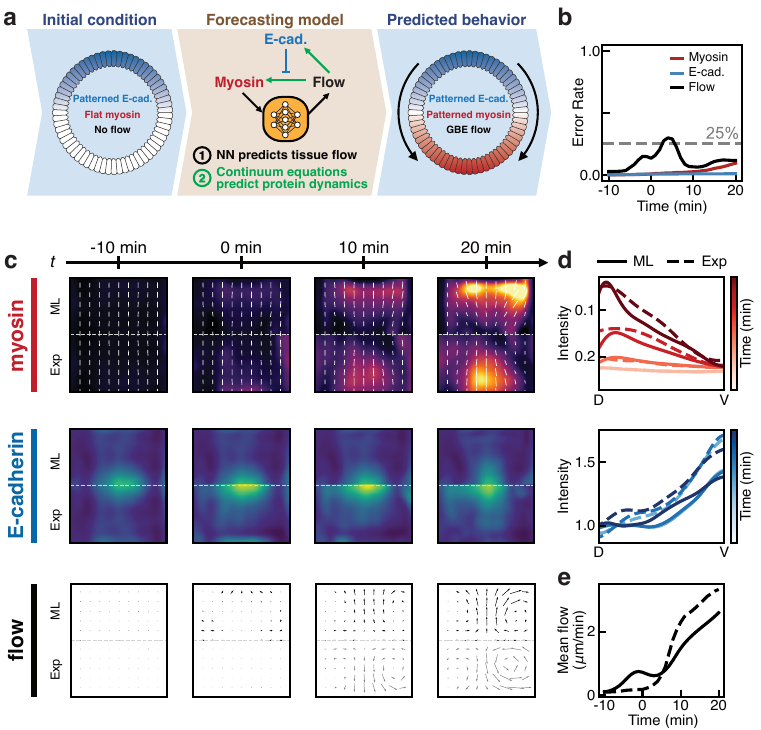}
    \caption{
        \textcolor{Highlight}{\textbf{Predicting dynamics of cytoskeletal proteins and flow. }}
        (\textbf{a}) Schematic of machine-learned dynamical system. At each time step, a NN (Fig.~\ref{fig:deepNN}f) predicts instantaneous tissue flow from myosin. Eqs.~\ref{eq:cadherin_sindy}-\ref{eq:myosin_sindy} (see Box 2) predict the time derivatives of the myosin and E-cadherin fields, which are integrated to forecast embryo behavior from initial conditions.
        (\textbf{b}) Error rate of predictions over 30 minutes, beginning 10 minutes before VF. The error for each field remains within $\approx$25\% over the trajectory.
        (\textbf{c}) Snapshots of predicted myosin, E-cadherin, and tissue flow at 10-minute increments. The top half of each row shows the model prediction (ML) while the bottom half shows the coarse-grained fields measured in experiment (Exp).
        (\textbf{d}) DV variation in model predictions (solid line) and experiments (dashed line) over time for myosin and E-cadherin. Myosin develops a DV gradient from uniform initial conditions, while E-cadherin maintains an initial DV gradient.
        (\textbf{e}) Average flow magnitude predicted by the model (solid line) and measured in experiment (dashed line). The model predicts the onset of GBE flow arising from the developing DV contractility gradient.
        }
    \label{fig:sindy}
\end{figure}

To evaluate these dynamical equations, we must also model the instantaneous relation between myosin and flow~\cite{streichanGlobalMorphogeneticFlow2018}.
Here, we use a neural network in order to account for external forces due to the ventral furrow invagination, another part of \emph{Drosophila} morphogenesis that we do not analyze (Fig.~\ref{fig:sindy}a-b).
Our hybrid model predicts the development of a dorso-ventral myosin gradient opposite to the E-cadherin gradient (Fig.~\ref{fig:sindy}c-e) and the transition to a vortical flow pattern (Fig.~\ref{fig:sindy}c).
We accurately forecast the developmental trajectory as a closed loop for up to 30 minutes, starting 10 minutes before initiation of the ventral furrow, and ending during GBE (Fig.~\ref{fig:sindy}b).
How does the anisotropy of the myosin nematic field emerge? Based on our model, this anisotropy originates from two sources: first, the geometry of the egg explicitly breaks the symmetry (this is modeled by the geometric stress term in Box 2); second, the formation of the ventral furrow pulls on the embryo (leading to the active tension and strain-rate recruitement in Box 2).
% TODO: add sausage explanation of geometric stress in Box 2
% TODO: map terms and pics in Box 2

\begin{cellbox}{boxbordercolor}{Box 2. Learning an interpretable model of embryo development}
    \footnotesize
    After considering minimal models, we use the SINDy method to learn rules from the data itself. In brief, this method constructs dynamic equations from a large \textit{library} of possible terms, with the aim of fitting the data with as few terms as possible (see SI for details). 
Eq.~\ref{eq:cadherin_sindy} is a passive advection equation for E-cadherin.
The cartoons below Eq.~\ref{eq:myosin_sindy} are visual interpretations of each term. 
The LHS is constrained to the co-rotational derivative for the given velocity field. 
The first RHS term captures the propensity for myosin motors to detach from junctions. 
The second is tension recruitment due to local myosin-driven active stress and the third is an embryo-scale static stress. 
Finally, there is a mechanical feedback which recruits myosin to strained junctions. \\

\begin{minipage}{0.98\textwidth}
    \includegraphics[width=1.02\textwidth]{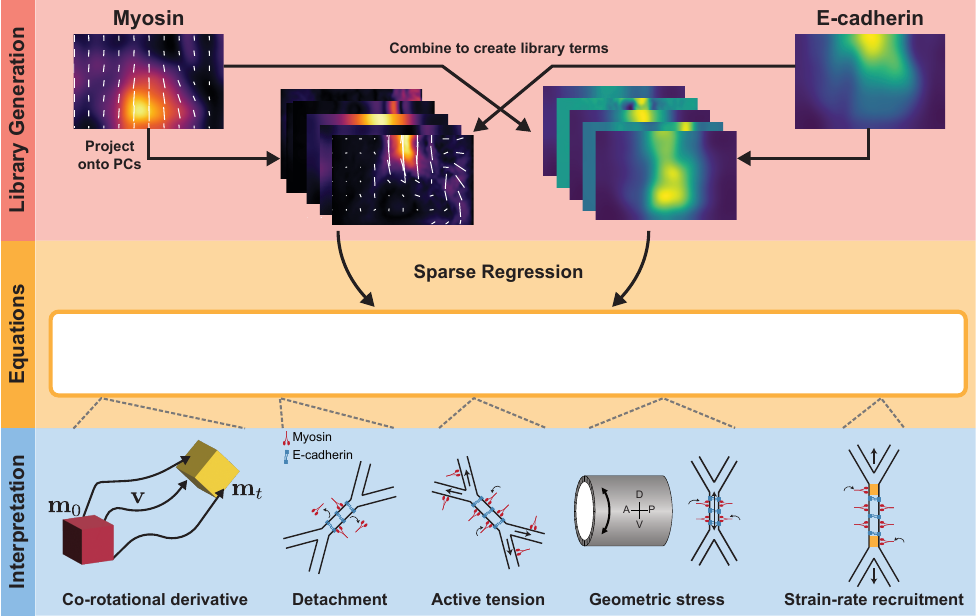}
    \vspace{-2.24in}
    \small
    \begin{equation}        
        \quad\ \partial_t c + (\mathbf{v} \cdot \nabla) c = 0
        \label{eq:cadherin_sindy}
    \end{equation}
    \begin{equation}
        D_t \mathbf{m} =
        -k_1 (1 - a_1 c) \ \mathbf{m} + 
        k_2 (1 - a_2 c) \ m\, \mathbf{m} + 
        k_{\Gamma} (1 - a_{\Gamma} c) \ m\, \bm{\Gamma}^{\text{DV}} + 
        k_E (1 + a_E c) \ E\, \mathbf{m}
        \label{eq:myosin_sindy}
    \end{equation}
    \vspace{1.4in}
\end{minipage}
The local density of cadherin (shown in blue in each junctional cartoon) modulates the strength of each feedback mechanism. For a cadherin pattern $c$, the
%Each mechanism is modulated by the cadherin pattern $c$. The 
$(1 - c)$ prefactors establish a myosin DV-gradient opposite cadherin.
The strain-rate recruitment term increases myosin concentration as a tissue is stretched.
We note that the coarse-grained field $\mathbf{E}$ cannot distinguish contributions from two distinct biological processes, cell rearrangement and cell stretching \cite{blanchardTissueTectonicsMorphogenetic2009}. Biologically, only the latter should recruit myosin. The learned strain rate term has a $(1+c)$ prefactor, which we interpret as distinguishing the strains from these two sources. Rearrangement dominates laterally while deformation primarily occurs dorsally, correlating with the cadherin pattern. Thus, the macroscopic strain rate term in (\ref{eq:myosin_sindy}) plays a different role than the microscopic (i.e. junctional) strain rate feedback measured in~\cite{gustafsonPatternedMechanicalFeedback2022, nishizawa_two-point_2023}.

\end{cellbox}

\clearpage
\subsection{Mutant analysis validates machine-learned rules: Tissue flow requires an embryo-scale adhesion pattern}
%Tissue flow requires cytoskeletal DV gradient downstream of BMP/DPP signalling.}

Our model (Box 2) predicts that tissue flow during GBE requires an embryo-scale gradient in cell adhesion (interpreted as the measured E-cadherin profile). 
We test this prediction using maternal mutants of  \textit{Drosophila} in which the DV patterning system is affected (Fig.~\ref{fig:mutants}).
We compare the wild type (first row of the figure) to two mutants known as \textit{spz$^4$} (second row) and \textit{Toll$^{RM9}$} (third row) \cite{stathopoulos_whole-genome_2002}.
These mutants adopt uniform cell fates corresponding to different positions along the DV axis of the wild-type embryo (arrows in first column of Fig.~\ref{fig:mutants}). 
Hence, the adhesion patterns are uniform along the DV axis (second column and panel b). 
Based on our machine-learned rules, we predict that no contractility gradient will form and no flow will occur in these mutants.
Indeed, we observe that the myosin DV gradient is strongly suppressed in the mutants (third column and Fig. 4c), and no significant tissue flow is observed (fourth column and Fig. 4d).
In both WT and mutant embryos, we observe an anti-correlation between E-cadherin and myosin levels in agreement with our learned rules. 
We note that in \textit{Toll$^{RM9}$} embryos, no flow occurs despite a uniformly high level of myosin, supporting the hypothesis that junctional myosin recruitment alone is not sufficient to drive tissue flow during GBE~\cite{streichanGlobalMorphogeneticFlow2018}. 
%In addition, we observe an anti-correlation between the uniform levels of E-cadherin and myosin in the mutants (compare second and third rows), in agreement with the learned rules.

%We characterize tissue flow, and cytoskeletal expression profiles in both lateralized (\textit{Toll$^{RM9}$}) and dorsalized (\textit{spz$^4$}) embryos, in which cells uniformly adopt lateral, neural ectodermal fates or dorsal ectoderm fates respectively. 
%Myosin recruitment is uniformly low in dorsalized embryos, and E-cadherin expression is uniformly high (Fig.~\ref{fig:mutants}a'-c',e-f).  In contrast, myosin recruitment is uniformly high in lateralized embryos whereas E-cadherin expression is low (Fig.~\ref{fig:mutants}a''-c'',e-f).  
%In both cases, graded cytoskeletal expression is eliminated (compare to Fig.~\ref{fig:mutants}a-c).  Importantly, tissue flow is nearly eliminated in both DV patterning mutant classes (Fig.~\ref{fig:mutants}g). 
%These results demonstrate that junctional myosin recruitment is not sufficient to drive convergent extension flow during GBE. 
%Indeed, lateralized embryos have high levels of myosin recruitment uniformly.  Rather, the generation of a contractility gradient is necessary for flow.

\subsection{Zygotic regulation: BMP/Dpp signaling controls 
%the post-translational machinery by patterning the 
embryo-scale adhesion distribution}

\begin{figure}
    \centering
    \includegraphics[width=0.82\textwidth]{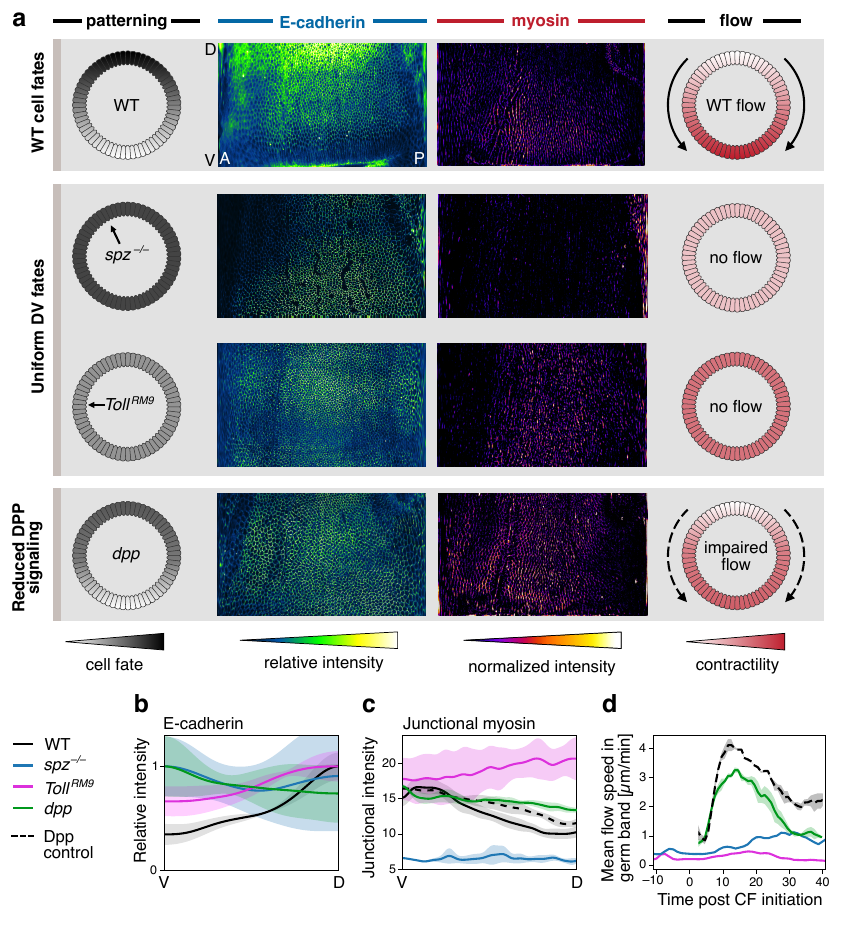}
    \caption{
        \textbf{\textcolor{Highlight}{DV patterning mutants suppress cytoskeletal gradients and tissue flow.}}
        (\textbf{a}) (first column) DV-patterning "clock" cartoon of \textit{Drosophila} gastrula. Color indicates differentiated cell fate.
        \emph{Toll$^{RM9}$} and \emph{spz$^4$} embryos adopt uniform fates corresponding to positions marked with arrows.
        \textit{dpp} mutants eliminate dorsal-most fates and lateral fates shift towards the dorsal pole.
        (second column) E-cadherin measured at initiation of the cephallic furrow. Fourth row shows E-cadherin in \emph{dpp$^4$ , snail$^{IIG05}$} double mutants.
        (third column) Cytosolic-normalized myosin measured 15 min post initiation of the cephallic furrow. Fourth row shows myosin measured in \emph{Dpp$^{hin46/+}$} mutants.
        (fourth column) Our model predicts that elimination of DV-patterned adhesion should impair or eliminate contractility gradients and tissue flow during GBE.
        (\textbf{b}) Quantification of E-cadherin gradient along DV axis for each genotype in (\textbf{a}).
        (\textbf{c}) Quantification of junctional myosin along DV axis for each genotype in (\textbf{a}). \emph{Dpp$^{hin46/+}$} shows impaired grading. Data in (\textbf{b-c}) based on $N=3$ embryos for each genotype. 
        (\textbf{d}) Quantification of tissue flow in each genotype. Flow is suppressed in \emph{Toll$^{RM9}$} and \emph{spz$^4$} mutants (imaged at 22$^\circ$ C). \emph{Dpp$^{hin46/+}$} mutants impair but do not abolish tissue flow. \emph{Dpp$^{hin46/+}$} and \emph{Dpp$^{+/+}$} control embryos were imaged at 27$^\circ$ C which increases the speed of gastrulation over embryos imaged at lower temperatures \cite{mitchellMorphodynamicAtlasDrosophila2022}.
    }
    \label{fig:mutants}
\end{figure}

At this stage, our tentative picture is entirely post-translational.
We now take a step back and ask: how does the embryo control the feedback loops involving only myosin and adherens junctions that we have identified?
The embryo-scale DV gradient of E-cadherin is established before the onset of gastrulation, likely by a DV-graded morphogen.
A natural candidate is Dpp, \textit{Drosophila} homolog of BMP, a key morphogen responsible for differentiation of cell fate along the DV axis~\cite{Padgett.etal1987}. 
Dpp is graded like E-cadherin with highest concentration at the dorsal pole. Therefore, it could control the E-cadherin concentration early on.
Reducing BMP signaling should then suppress the E-cadherin DV gradient, leading to a weaker contractility gradient and tissue flow.
To test this, we analyzed \textit{dpp$^{hin46/+}$} mutants (Fig.~\ref{fig:mutants}, fourth row), in which BMP signaling is reduced. We observe flattened E-cadherin and myosin patterns, as well as weaker flow relative to control embryos.
%Having identified the upstream role of BMP signaling in setting up the embryo-scale adhesion pattern, we now ask what is the signaling pathway by which this takes place.
This supports our hypothesis that BMP signaling establishes a DV-graded profile of E-cadherin and other components of the adherens junction.

\begin{figure}
    \centering
    \includegraphics[width=\textwidth]{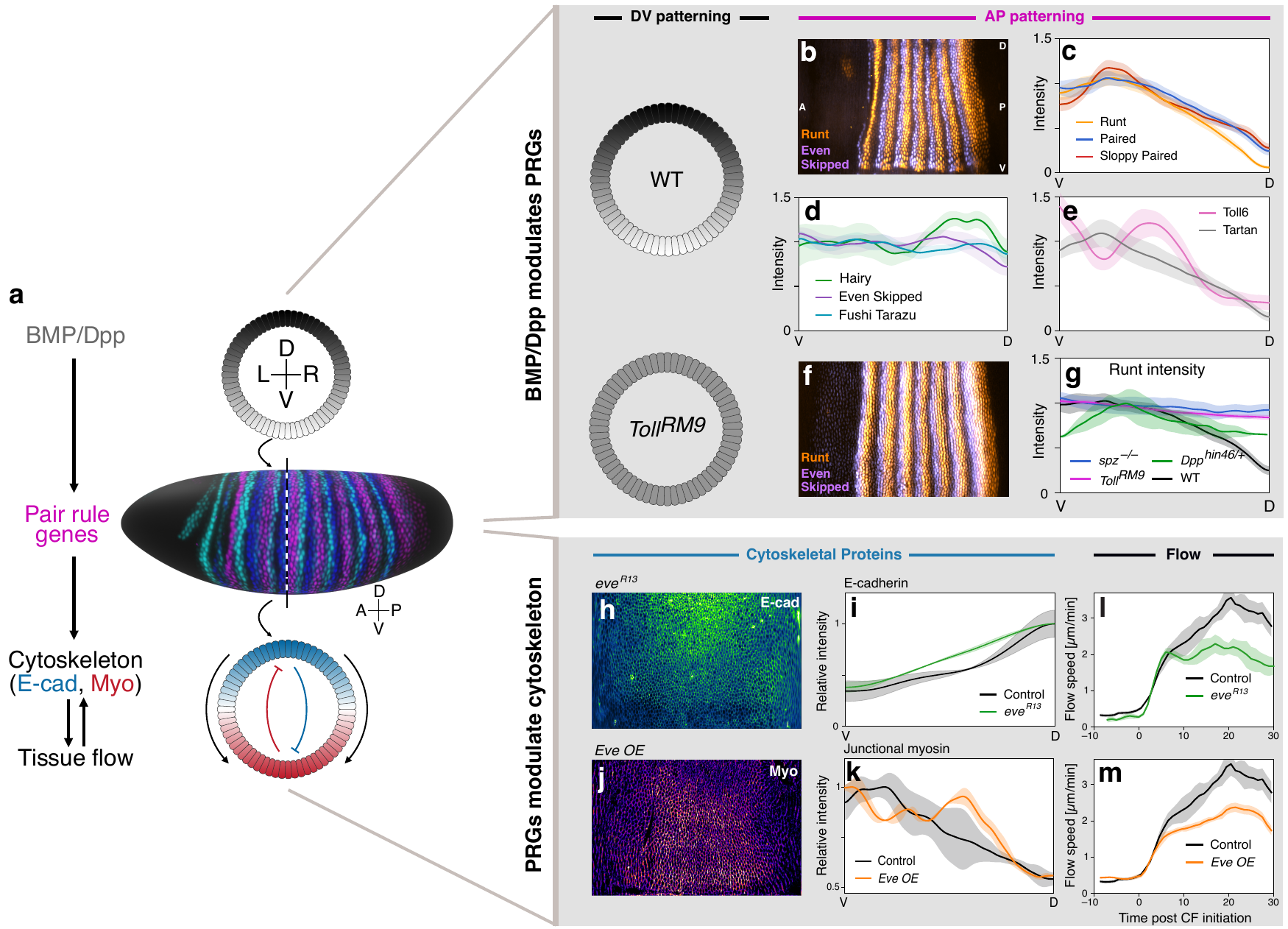}
    \caption{
    \textcolor{Highlight}{\textbf{BMP signaling controls protein behavior via PRGs.}}
        (\textbf{a}) Cartoon of proposed signaling cascade.
        (\textbf{b-e}) In addition to stripes along the AP axis, the expression of certain PRGs exhibits a modulation along the DV axis.
        PRG expression along a central slice shows DV modulation for Runt, Paired, and Sloppy Paired (c) and weak modulation for Hairy, Even Skipped, and Fushi Tarazu (d). Downstream transmembrane receptors (Toll-6, Tartan) are also DV-graded (e).
        %\jc{I think this plot should be re-done to show Runt and Paired, or Runt, Paired, and Even-Skipped, since these are the only ones present in the figure. Remaining PRGs are relegated to SI. Not sure on whether Tolls should be kept or relegated to SI}
        (\textbf{f}) \emph{Toll$^{RM9}$} mutants adopt uniform fates along the DV axis, but still exhibit AP stripes. Runt expression in these mutants shows no DV gradient.
        (\textbf{g}) Quantified Runt expression in DV mutants (Fig.~\ref{fig:mutants}) shows impaired or suppressed DV-modulation of myosin and E-cadherin.
        (\textbf{h-i}) \textit{eve}$^{R13}$ is a null allele of Even-skipped. E-cadherin expression in these mutants exhibits a shallower DV-gradient compared to control ($N=3$ embryos).
        (\textbf{j-k}) \textit{Eve OE} (\textit{67,15$\gg$UAS-eve}) over-expresses Even-Skipped uniformly throughout the embryo. Myosin expression in \textit{Eve OE} mutants is stronger near the dorsal pole of the embryo ($N=2$ embryos).
        (\textbf{l-m}) 
        Mutations to a PRG (\textit{eve$^{R13}$} and \textit{Eve OE} lead to weaker GBE flow ($N=3$ embryos per genotype).
        %Together, these results suggest that PRG mutations impact embryo-scale adhesion and cytoskeletal patterning, which in turn impair tissue flow. % already in text
    }
    \label{fig:patterning}
\end{figure}

%\subsection{BMP/Dpp signaling may act via PRGs to drive graded cytoskeletal recruitment.}
\subsection{Signaling pathway linking AP to DV axes: BMP/Dpp role in patterning adhesion via PRGs}

%We have shown that BMP signaling establishes a DV-graded profile of E-cadherin and other components of the adherens junction. 
%We now ask: how is this profile established (Fig.~\ref{fig:patterning})?
We now ask: what is the signaling pathway by which BMP establishes the embryo-scale adhesion pattern?
Previous work has established that PRGs are involved in GBE (Box 1), often focusing on their prominent striped AP dependence \cite{Zallen2004,irvineCellIntercalationDrosophila1994,Pare2014}.
Looking more closely, we observe that the expression levels of certain PRGs also exhibit gradients in the DV direction (Fig.~\ref{fig:patterning}b-d).
Note also that Toll-6, a downstream target of the PRG Even-Skipped, does exhibit a strong DV dependence in WT embryos (Fig.~\ref{fig:patterning}e), even though the expression of Even-Skipped itself does not show a strong DV dependence (Fig.~\ref{fig:patterning}d).
This motivates us to propose that BMP signaling may regulate E-cadherin through PRGs.

We test this hypothesis in two stages. To evaluate whether BMP/Dpp regulates PRG expression, we compare the PRG Runt
%To test this hypothesis, we first compare the PRG Runt 
in a WT embryo to the mutants \emph{Toll$^{RM9}$} and \emph{spz$^4$} that we introduced in Fig.~\ref{fig:mutants}. While the AP-oriented stripes are unchanged, we observe that the DV gradient of Runt present in the WT disappears in the mutants (Fig.~\ref{fig:patterning}f-g).

To test whether PRGs regulate E-cadherin,
we then turn to mutants where PRG expression is explicitly modified (see Methods). %, focusing on Even-Skipped. 
We observe that both inhibition of Even-Skipped via \emph{eve$^{R13}$} mutants and upregulation via \emph{67,15}$\gg$\emph{UAS-eve} exhibit a weaker tissue flow compared to control embryos (Fig.~\ref{fig:patterning}l-m).
In addition in \emph{eve$^{R13}$} mutants, we observe reduction in the levels of E-cadherin in the middle of the DV axis (Fig.~\ref{fig:patterning}h-i). Consistently over-expression via \emph{67,15}$\gg$\emph{UAS-eve} shows an extension of the myosin pattern to the dorsal pole (Fig.~\ref{fig:patterning}j-k).

Together, these observations support the hypothesis that BMP signaling acts at least in part via the AP-patterned PRGs to establish a smooth DV-graded profile of E-cadherin expression before GBE. 
This gradient in cell adhesion then leads to a gradient in myosin recruitment along the DV axis, possibly by modulating mechanical feedback loops (see Box 2, Refs.~\cite{gustafsonPatternedMechanicalFeedback2022,Lefebvre.etal2022}, and SI). The resulting contractility gradient drives tissue flow. 
A visual summary of the proposed mechanism is show in Fig.~\ref{fig:patterning}a.

\begin{figure}
    \centering
    \includegraphics[width=0.75\textwidth]{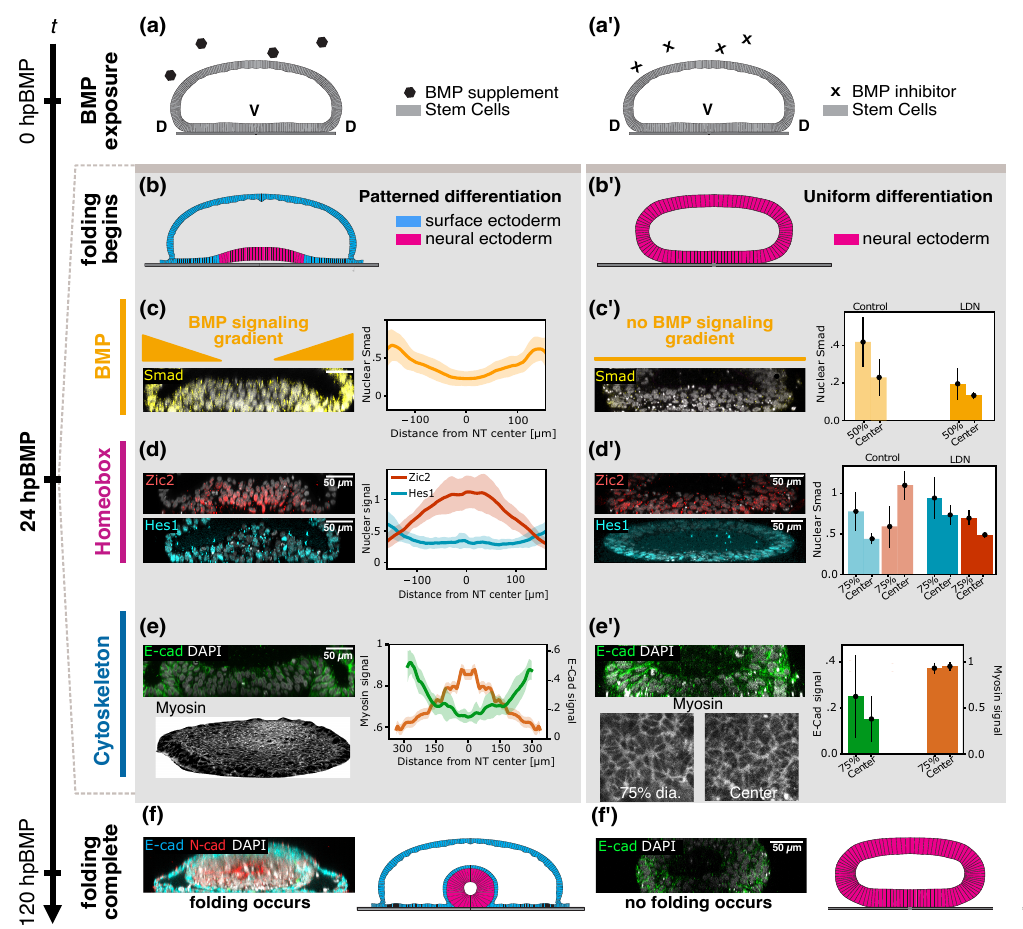}
    \caption{\textcolor{Highlight}{\textbf{Patterned expression in human neural tube organoids (NTO). }}
        (\textbf{a}) 3D stem cell sheets are exposed to BMP4 to drive tissue folding. All times are measured as hours post-BMP exposure (hpBMP).
      (\textbf{b}) By $t = 24$ hpBMP, the tissue differentiates into surface and neural ectoderm and first morphogenetic movements begin. (\textbf{c}) Left: Nuclear SMAD 1/5/9 (Smad), a BMP4 signaling marker, is used to readout BMP4 signaling. NTO stained for Smad shows an expression gradient, taken at $t=20$ hpBMP.  Right: Quantification obtained from the average nuclear signal across $z$-slices (surface ectoderm dome excluded) in $N=6$ NTOs.
      (\textbf{d}) Left: Zic2 and Hes1 are homologs of \emph{Drosophila} PRGs. NTOs stained for Zic2 and Hes1 exhibit expression gradients in relation to the SMAD gradient, taken at $t=20$ hpBMP.  Right: Quantification obtained using $N=6$ NTOs.
      (\textbf{e}) Left: E-cadherin, taken at $t=20$ hpBMP, and apical myosin, taken at $t=24$ hpBMP, expression in NTOs. Both exhibit a gradient in relation to Smad.  Right: Quantification obtained using $N=6$ NTOs for E-cadherin, and $N=2$ NTOs for myosin.
      (\textbf{f}) The folding process is complete by $t = 120$ hpBMP.
      (\textbf{a'-f'}) Repeated analysis on NTOs treated with BMP4 inhibitor LDN (200\textit{n}M).
      The resulting NTO lacks surface ectodermal differentiation (\textbf{b'}) as signified by absence of Smad straining (\textbf{c’}, Left).  Right: Quantification of Smad staining in two different positions along the NTO DV axis in structures treated with LDN ($N=6$ NTOs, right) and in control NTOs ($N=6$ NTOs, left).  (\textbf{d’}) Left: In LDN treated structures, homebox gene expression profiles exhibit no significant gradients.  Right: Quantification of homeobox gradients in control NTOs (Hes1: light blue, $N=8$ NTOs, Zic2: light red, $N=10$ NTOs) and LDN treated NTOs (Hes1: dark blue, $N=9$ NTOs, Zic2: dark red, $N=5$ NTOs) at two different positions along the NTO DV axis. (\textbf{e’}) Left: Gradients of cytoskeletal components (myosin: gray, E-cadherin: green) are gone in NTOs treated with LDN.  Right: Quantification of E-cadherin ($N=5$ NTOs) and Myosin ($N=4$ NTOs) expression at two different (75\% and center) radial positions along the NTO.
(\textbf{f’}) No folding occurs in structures that are treated with LDN.  The one-sided Welch t-test was used to test if the gradient (margin-center difference) is greater in control than in the LDN treated NTOs. P-values are as follows: Zic2: $1.5 * 10^{-5}$, Hes1: 0.16, Smad: 0.06, E-cadherin: 0.17, Myosin: 0.16.    
        %(\textbf{a}) 3D stem cell sheets are exposed to BMP4 to drive tissue folding. All times are measured as hours post-BMP exposure (hpBMP).
        %(\textbf{b}) By $t = 24$ hpBMP, the tissue differentiates into surface and neural ectoderm and first morphogenetic movements begin. (\textbf{c}) Left: Nuclear SMAD 1/5/9 (Smad), a BMP4 signaling marker, is used to readout  BMP4 signaling. NTO stained for Smad shows an expression gradient, taken at $t=20$ hpBMP.
        %Right: Quantification obtained from the average nuclear signal across $z$-slices in NTO (surface ectoderm dome excluded) in $N=6$ NTOs).
        %(\textbf{d}) Left: Zic2 and Hes1 are homologs of \emph{Drosophila} PRGs. NTOs stained for Zic2 and Hes1 exhibit  expression gradients in relation to the SMAD gradient , taken at $t=20$ hpBMP.
        %Right: Quantification obtained using $N=6$ NTOs.
        %(\textbf{e}) Left: E-cadherin, taken at $t=20$ hpBMP, and apical myosin , taken at $t=24$ hpBMP, expression in NTOs. Both exhibit a gradient in relation to Smad.
        %Right: Quantification obtained using $N=6$ NTOs for E-cad. and $N=2$ NTOs for myosin.
        %(\textbf{f}) The folding process is complete by $t = 120$ hpBMP.
        %(\textbf{a'-f'}) Repeated analysis on NTOs treated with BMP4 inhibitor LDN (200\textit{n}M).
        %The resulting NTO lacks surface ectodermal differentiation (\textbf{b'}) as signified by absense of Smad straining (\textbf{c'}), homebox and E-cadherin expression profiles exhibit no significant gradients (\textbf{d'-e'}), and no folding occurs (\textbf{f'}).
    }
    \label{fig:neural_tubes}
\end{figure}

\subsection{Conserved role of BMP and adhesion patterning from \textit{Drosophila}  to human neuroectoderm} %morphogenesis}
%BMP signaling has a conserved role in neuroectoderm fate determination \cite{bier_evolution_2011}. In \emph{Drosophila}, we have established a link between BMP signaling and a global smooth gradient of adherens junctions, that  via dynamic feedback establishes a counter running myosin gradient.  (Fig.~\ref{fig:patterning}a). We now ask: to what extent is the link between BMP signaling and proteins at the adherens junction conserved across species?
BMP signaling has a conserved role in neuroectoderm fate determination from fly to vertebrates~\cite{Bier2011}. Molecular components of the cytoskeleton, such as myosin or E-cadherin, are also conserved between the species. Yet the morphogenesis of these tissues involves distinct shapes in different species. 
%For instance, in \emph{Drosophila} the germband extends along the surface of an ellipsoid but does not fold, while in humans the neural tube folds on the surface of a disc. In \emph{Drosophila}, we established a link between BMP signaling and a global smooth gradient of adherens junctions that establishes a myosin gradient in the opposite direction via dynamic feedback (Fig.~\ref{fig:patterning}a). The model relies on mechanical factors as a key component to organize myosin patterns. The different initial shapes of the neuroectoderm are expected to have a profound impact on mechanics. 
This raises the question of whether the post-translational mechanism we have identified in \emph{Drosophila} (and summarized in Fig.~\ref{fig:patterning}a) can be conserved in organisms with different geometries. 

To address this question, we turn to a reproducible human stem cell-based model of neural tube morphogenesis. In this model, stem cell cultures with a single lumen, controlled geometry, volume, and cell number are differentiated into neuroectoderm by neural induction and timed BMP4 supplement (Fig 6a), see Methods and Refs.~\cite{Karzbrun.etal2021,Britton.etal2019}. 
In contrast with \emph{Drosophila}, where the germband extends along the surface of an ellipsoid but does not fold, in humans the neural tube folds on the surface of a disc.
BMP signalling is driven from the edge, and the resulting gradient organizes patterned differentiation into surface ectoderm (in blue in Fig 6b) and neural ectoderm (in purple in Fig 6b)~\cite{etoc_balance_2016}. In this stem cell-based model, the neural tube undergoes major shape changes, including bending, folding and closure~\cite{Karzbrun.etal2021}. 
These major shape changes have important consequences on mechanics, and may complicate the interpretations of cytoskeletal patterns.
Hence, we focus on the early configuration of the cytoskeleton, that triggers the flows leading to bending (i.e. before 36 hours post BMP supplementation (hpBMP), see Methods). 
%In this model, stem cell cultures with a single lumen, and  controlled geometry, volume, and cell number are differentiated into neuroectoderm by neural induction and timed BMP4 supplement (Fig 6a).
%BMP signalling is driven from the edge, such that a graded profile arises. This BMP signaling gradient organizes patterned differentiation into  surface ectoderm (in blue in Fig 6b) andneural ectoderm (in purple in Fig 6b). 
%The neural tube eventually folds leading to a closure (FIG).

% FIGURE NEEDS ROSETTA STONE DROSOPHILA/HUMAN
We find that all the elements present in our proposed mechanism for \textit{Drosophila} (Fig.~\ref{fig:patterning}a) also show organ-scale gradients in the stem cell-based model, and mirror BMP signaling. By 20 hpBMP we observe a stable BMP signaling gradient (Fig.~\ref{fig:neural_tubes}c). At the same time, we see differential expression patterns of homeobox genes such as Zic2 and Hes1, homologs of \textit{Drosophila} PRGs (Fig.~\ref{fig:neural_tubes}d). Adherens junction proteins such as E-cadherin are also smoothly graded at this time, and only become sharply delineated from the neural ectoderm later. By 24 hpBMP, myosin exhibits a smooth gradient that counter runs the E-cadherin profile (Fig.~\ref{fig:neural_tubes}e). 
We note that because the geometry of the human \textit{in vitro} neural tube differs from that of \textit{Drosophila}, there is no geometric stress (Box 2) to drive local anisotropic recruitment of myosin. Nevertheless, the comparison with \emph{Drosophila} suggests that both force-generating proteins and PRG orthologs are modulated by the BMP signaling gradient, suggesting that this role of BMP in setting the initial condition for the morphogenetic events is conserved. 

To test this further, we eliminate BMP signaling via the BMP inhibitor LDN (Fig~\ref{fig:neural_tubes}a'), so that there is no BMP gradient (Fig~\ref{fig:neural_tubes}c'). As expected, all cells differentiate into neural ectoderm: the resulting structures completely lack surface ectoderm (compare Fig~\ref{fig:neural_tubes}b and b'). We further observe that, as a consequence, PRG orthologs and E-cadherin also exhibit no gradient (Fig. 6d',e'). Finally, the structures where BMP signaling is suppressed do not fold (Fig 6f'). This supports our hypothesis that the organoid-scale gradient in BMP signaling primes the cytoskeleton to orchestrate the morphological movements that initiate neural tube closure, possibly through PRG homologs. 

To sum up, despite striking differences in shape, the \textit{in vitro} neural tube model and \emph{Drosophila} share a crucial geometrical feature: there is a key axis defined by BMP patterning in both systems, along which a contractility gradient gets established. Our results suggest that this general feature is conserved, but fine-tuned by the geometry of different organisms.

\section{Discussion}

\begin{figure}
    \centering
    \includegraphics[width=0.8\textwidth]{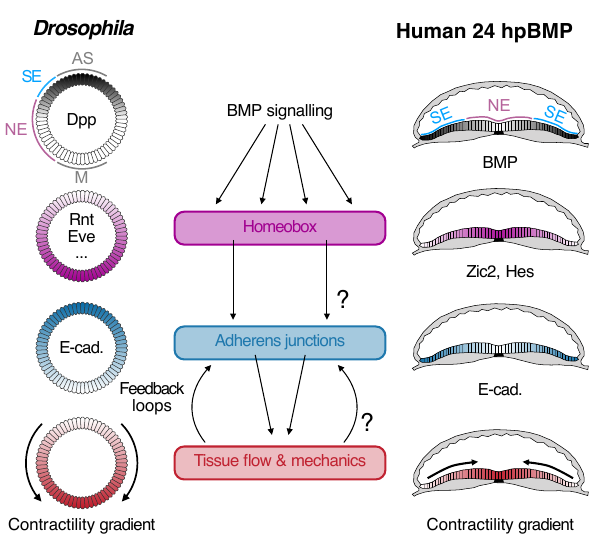}
    \caption{\textcolor{Highlight}{\textbf{Proposed conserved signaling cascade from BMP signaling to tissue flow.}}
    %DV patterning from the fly to the human neural tube.
    Hypothesized morphogenetic parallels between \textit{Drosophila} gastrulation and human neural tubes.     SE: surface ectoderm, NE: neural ectoderm, M: mesoderm, AS: amnioserosa.
    DPP/BMP signaling establishes a gradient in Homeobox transcription factors (PRGs in \textit{Drosophila}), which lead to large scale cytoskeletal gradients, notably of E-cadherin and adherens-junctional proteins. These gradients lead to tissue flow, which can feed back to the cytoskeleton.
    Open questions in the neural tube organoids are the causal role of homeobox gradients, and the presence or absence of mechanical feedback loops.
    %\textcolor{red}{is the E-cadherin gradient in the wrong direction? at least it is not the same as in previous figure (or I don't get the color code)} 
  }
    \label{fig:graphical_abstract}
\end{figure}

In this work, we have proposed a global picture of neuroectoderm morphogenesis which links the conserved mechanisms of BMP signaling and convergent extension. 
%We have argued that this overall mechanism too is conserved across species. 
Figure~\ref{fig:graphical_abstract} summarizes how this picture materializes in  \emph{Drosophila} neuroectoderm morphogenesis (left) and in a human neural tube model (right).
In both cases, BMP signaling gradients are relayed into a global pattern in adherens junction proteins via genetic regulation. 
This in turn sets up a global pattern of force-generating cytoskeletal proteins, such as myosin, that modulate tissue mechanics.
In \emph{Drosophila}, this arises through mechanical feedback loops involving tissue-scale stress and external strains, which result in coordinated morphogenetic flow~\cite{gustafsonPatternedMechanicalFeedback2022,lefebvre_geometric_2023}.
%This results in a coordinated morphogenetic flow during axis elongation in \emph{Drosophila}. 
Whether human neural tube closure relies on similar feedback mechanisms for governing protein dynamics remains an open question, but we note that Zic family proteins are involved in neural tube morphogenesis ~\cite{Nagai.etal2000}.
%A key feature of our mechanism is the tight integration between dorso-ventral (DV) and anterior-posterior (AP) patterning systems. 
This overall bio-mechanical picture might be conserved even beyond bilaterians. 
In \emph{Nematostella} (sea anemones), for example, BMP signaling regulates a distinct set of homeobox genes \cite{Genikhovich.etal2015} also associated with the mechanical processes sculpting the body \cite{He.etal2018}.

The picture we have developed aligns with classical work showing that inverting the BMP gradient via RNA injection reverses the direction of \emph{Drosophila} neuroectoderm extension~\cite{Ferguson.Anderson1992}.
%In our model, BMP signaling provides the critical ingredient by breaking the DV symmetry of the embryo.
However, it is complementary to the {heterotypic juxtaposition model}~\cite{Pare.Zallen2020}. 
%of \emph{Drosophila} neuroectoderm extension
Instead of direct genetic regulation locally modulating %cell adhesion and 
anisotropic forces at cell boundaries,
our model emphasizes the role of a embryo-scale modulation of adherens junctions which breaks the DV symmetry in a manner prescribed by BMP signalling. Concurrent work has reported a similar global DV patterning of adherens junction proteins other than E-cadherin \cite{clarke_efgrdependent_2023}. 
A molecular mechanism for this large-scale modulation of adherens junction proteins may involve Toll-like transmembrane receptors (TLRs; see Fig.~\ref{fig:patterning}b).
%, which were recently discovered in a broad class of invertebrate systems \cite{Benton.etal2016}%, could indicate a molecular mechanism for this large-scale modulation of adherens junction proteins 
%().
The precise mechanism by which TLRs regulate the cytoskeleton remains incompletely understood \cite{Pare.Zallen2020}. However, TLRs are known to strongly impact neural epithelium elongation in \emph{Drosophila} and to be controlled by PRGs~\cite{Lavalou.etal2021,Tamada.etal2021}. 
%Yet how precisely TLRs regulate the cytoskeleton remains incompletely understood \cite{Pare.Zallen2020}. 
%Indeed, Fig.~\ref{fig:patterning}b suggests a global DV modulation patterns in TLRs. 

The picture of neuroectoderm morphogenesis proposed in this work has been obtained by developing a set of causal rules, that we then tested in experiments.
These causal rules were identified by complementing data-driven inference with insights from biology and physics.
This is not a new way of doing biology. 
Nonetheless, our use of machine learning techniques aids the formulation and testing of biological hypothesis and mathematical models~\cite{Schmitt.etal2023}. 
%Complementing machine learning with insights from biology and physics softened the trade-off between interpretability and fidelity to experiments, enabling us to learn interpretable and predictive rules.
%Ultimately, we identified interpretable rules only after complementing machine learning with insights from biology and physics. 
What we can learn also depends on the interpretative framework that we use. 
For instance, our coarse-grained picture neglects the details of cell-level mechanisms. As a result, our model shows discrepancies with experimental measurements of mechanical feedback at strained junctions (Box 2 and SI), which are relevant during ventral furrow invagination~\cite{gustafsonPatternedMechanicalFeedback2022}. 

We conclude by asking: why should the mechanisms by which morphogenesis unfolds be conserved? 
In biology, usual sources of universality are the conservation of genes by natural selection and convergent evolution.
Morphogenesis has the peculiarity of strongly mixing biological and mechanical constraints: organisms have to abide to the laws of mechanics to get their shape.
We conjecture that these physical constraints translate into evolutionary constraints that make morphogenesis more likely to exhibit conserved mechanisms.
%These constraints come together in the cytoskeleton, where genetics codes for conserved proteins that in turn execute the mechanical processes shaping tissues. 
Our approach may accelerate the systematic identification of such universal mechanisms -- conserved across species -- that biology uses to seed the shape of organisms.

\clearpage

\bibliographystyle{unsrt.bst}
\bibliography{ML_Drosophila.bib}

\paragraph{Acknowledgments} We thank Susan Wopat, Aimal Khankhel, and the members of the Streichan lab for helpful discussions. 
This work was supported by grants NIGMS R35-GM138203 and CAREER PHY:2047140. 
JC and MF acknowledge support from the National Science Foundation under grant DMR-2118415.
MF and VV acknowledge partial support from the UChicago Materials Research Science and Engineering Center (NSF DMR-2011864).
VV acknowledges support from the Army Research Office under grant  W911NF-22-2-0109 and W911NF-23-1-0212 and the Theory in Biology program of the Chan Zuckerberg Initiative. 
This research was supported from the National Science Foundation through the Center for Living Systems (grant no. 2317138).
This work was completed in part with resources provided by the University of Chicago’s Research Computing Center.

\paragraph{Declaration of interests} The authors declare no competing interests.

%\paragraph{Author contributions} \nc{To do for cell} 

%\begin{comment}

\newpage

\clearpage

\section{Materials and methods}

\subsection{\emph{Drosophila} stocks}
Lateralized embryos were generated by collecting embryos from \textit{Klar, Tl$^{RM9}$ Sqh::GFP / Klar, Tl$^{RM9}$ Sqh::GFP} homozygous females.  Dorsalized embryos were generated by collecting embryos from trans-heterozygous \textit{Klar, Spz$^1$ Sqh::GFP / Klar, Spz$^4$ Sqh::GFP} females. Even-Skipped homozygous mutant embryos, live imaged using a single maternal copy of the sqh::GFP transgene were generated by crossing \textit{$\Delta$J29 Even-Skipped$^{R13}$ /Cyo, Sqh::GFP} males and females. 
The visible phenotype halo \textit{($\Delta$J29)} was used to identify homozygotic mutant embryos during early cellularization. Even-Skipped over expression embryos were generated via the following crosses.  
Virgin females, homozygous for ubiquitous maternal Gal4 drivers on both the second and third chromosomes (\textit{67/67;15/15} see Supplementary Table 1), were crossed to \textit{+/+ ; +/Cyo , Sqh::GFP} males.  Virgin female progeny (\textit{67/+ ; 15/Cyo, Sqh::GFP}) were then crossed to \textit{+/+ ; UAS-Even-Skipped/UAS-Even-Skipped} males. 
The resulting progeny had the \textit{UAS-eve} transgene driven by two maternal GAL-4 drivers and were visualized using \textit{Sqh::GFP}.  To image myosin expression live in a Dpp mutant background, we used the haploinsufficient null \textit{Dpp$^{hin46}$} allele.  We crossed \textit{dpp+/Cyo, Sqh::GFP} virgin females to \textit{Dpp$^{hin46}$ / Cyo, Dpp+} males (see Supplementary Table 1).  Both \textit{+/Dpp$^{hin46}$} and \textit{Dpp$^{hin46}$/Cyo, Sqh::GFP} were lethal progeny and produced a strong phenotype during GBE. 
These were compared to 'control' embryos which had the \textit{Cyo, dpp+} duplicated chromosome and therefore had two copies of WT dpp.  E-cadherin expression was analyzed in fixed embryos in \textit{dpp$^4$, snail$^{IIG05}$} double mutants. Male and female \textit{$\Delta$J29 dpp$^4$, snail$^{IIG05}$/Cyo Sqh::GFP} flies and collecting embryos that had the visible halo \textit{($\Delta$J29)} phenotype at cellularization  for analysis. 
These embryos were imaged live using the \textit{Sqh::GFP} transgene.  The dpp$^4$ allele is a strong hypomorphic allele that is still heterozygous viable.  Fluorescent transgenes used for quantification in this study include \textit{Sqh::GFP}, and \textit{Endo-Ecadherin::GFP}.

\subsection{Human Embryonic Stem Cell Lines}
We used derivatives of the human embryonic stem cell lines RUES2 (Rockefeller) and ESI017 (ESI-Bio) in this study. The two variants of these parental lines used in this paper include RUES2 SMAD4::GFP and H2B::RFP \cite{nemashkaloMorphogenCommunityEffects2017} and ESI017 E-cadherin::GFP H2B::RFP \cite{masseySynergyTGFvLigands2019}. Maintenance culture conditions were done using mTeSR Plus (Stem Cell Technologies) as described previously \cite{Karzbrun.etal2021}.

\subsection{Generation of human neural tube organoids}

Organoid generation has been previously described \cite{Karzbrun.etal2021}.  Briefly,  human embryonic stem cells (see Supplementary Table 1) were maintained in mTeSR Plus (Stem Cell Technologies, 100-0276).  On day 0 of the organoid generation protocol cells were seeded on laminin (Biolamina, MX521-0501) coated micro-patterns on glass bottomed culture dishes and allowed to adhere overnight in mTeSR Plus.  On day 1, maintenance media was replaced with neural induction media supplemented with 5 $\mu$M TGF-$\beta$ inhibitor SB431542 (Stem Cell Technologies, 72232).  Neural induction media was also supplemented with 2\% HESC-qualified matrigel (Corning, CLS356277) to promote lumen formation.  On day 2, structures were left undisturbed to facilitate lumen formation.  Media was exchanged on day 3, and on day 4, neural induction media was exchanged and supplemented with 5 ng ml$^{-1}$ BMP4 as well as 5 $\mu$M SB431542.  To investigate the early events of neural tube formation, some structures were fixed after 24 hours of BMP4 exposure, while other structures were grown for an additional 72 hours in neural induction media supplemented with BMP4.  Media exchanges occurred every day.  For BMP4 knockdown experiments, the small molecule BMP inhibitor LDN193189 (Thermo Fisher Scientific, 04-0074-02) was added to the culture media every day starting at protocol day 0 at a concentration of 200nM, and no BMP4 was supplied.

\subsection{Immunohistochemistry}
Heat fixation of \textit{D. melanogaster }embryos was done as described previously \cite{mullerArmadilloBazookaStardust1996}.  Briefly, embryos were dechorionated in 50\% bleach, and fixed via 10 second plunge submersion in boiling saline solution immediately followed by cooling on ice.  Devitellinization was done using 1:1 heptane:methanol immersion.  For analysis of E-cadherin protein expression using immunofluorescence, \textit{D. melanogaster} embryos were fixed in 1:1 heptane: 4\% paraformaldehyde for 20 minutes followed by devitellinization using 1:1 heptane:methanol immersion.

For immunofluorescent stain of protein expression in human neural tube organoids, structures were fixed by immersion in 4\% paraformaldehyde for 1 hour followed by washing with PBS. 

A full list of primary and secondary antibodies used can be found in Supplementary Table 1.  \textit{D. Melanogaster} embryos were blocked for 15 minutes using Image-iT signal enhancer (Thermo Fisher Scientific, I36933) before overnight incubation with primary antibodies at 4 degrees celsius.  Secondary antibodies were incubated for 2 hours at room temperature.  Immunofluorescent staining of human neural tube organoids was done as previous described \cite{Karzbrun.etal2021}.

\subsection{Imaging}
Confocal fluorescent imaging of human neural tube organoid structures was done using a Nikon x40/1.3 NA water immersion lens on a Leica SP8 microscope.  Organoid structures were imaged in-situ on the glass bottomed dishes where they were cultured.  Fixed organoid structures were cleared prior to imaging using RapiClear 1.47 (SUNJin Lab).  Live organoid cultures were imaged in a culture chamber maintained at 37 degrees celsius with 5\% CO$_2$. 

Fluorescent light sheet microscopy was used to image both fixed and live \textit{D. Melanogaster} embryos as described previously \cite{lefebvre_geometric_2023,gustafsonPatternedMechanicalFeedback2022,streichanGlobalMorphogeneticFlow2018}.  Briefly, live embryos were dechorionated in bleach, and mounted in a column of 1.5\% low melting point agarose for imaging on a custom confocal \cite{medeirosConfocalMultiviewLightsheet2015} MuVI SPIM microscope using Nikon x25/1.1 NA immersion detection objectives \cite{krzicMultiviewLightsheetMicroscope2012}.  Embryos were imaged at 22 degrees Celsius while submerged in de-ionized H$_2$O with a time resolution of 1 minute.

\subsection{Image analysis, data fusion, and extraction}

Multiple images acquired from different angular perspectives of the same sample were registered to each other using the position of fiduciary fluorescent beads (Fluoresbrite multifluorescent 0.5 $\mu$m beads, Polysciences Inc, 24054) as described previously \cite{gustafsonPatternedMechanicalFeedback2022} using the Multiview Reconstruction plugin for FIJI \cite{preibischEfficientBayesianbasedMultiview2014}.  The all-to-all image registration option was used to map all angles across all timepoints to a common reference frame.  Deconvolution was done using the Multiview Reconstruction plugin \cite{preibischEfficientBayesianbasedMultiview2014}.  Resultant 3D volumetric datatsets had isotropic spatial resolution of 0.2619$\mu$m.  2D surfaces of interest — such as the sub-apical adherens junction plane — were extracted from full 3D volumetric data using the ImSAnE surface detection algorithm \cite{heemskerkTissueCartographyCompressing2015} in tandem with Ilastik image segmentation software.  Two sequential iterations of the Ilastik/ImSAnE workflow resulted in an volumetric representation of a surface of interest (SOI) with multiple ‘onion layers’ that were spaced 1.5$\mu$m apart.  Maximum intensity projections of 9 onion layers were used to create the final SOIs for cytoskeletal proteins while maximum intensity projections of 20 onion layers were used for nuclear proteins.

\clearpage

\section{Supplementary Information}
\setcounter{figure}{0}
\renewcommand{\figurename}{Fig.}
\renewcommand{\thefigure}{S\arabic{figure}}

\subsection{Data handling: image processing and ensemble time alignment}

\subsubsection{Image processing}

To obtain coarse-grained data for training the deep NNs and sparse dynamical equations, we used the morphodynamic atlas of \textit{Drosophila} development~\cite{mitchellMorphodynamicAtlasDrosophila2022}.
This contains time series' of fluorescent images of a variety of biomarkers including cytoskeletal proteins and PRG expression patterns.
To convert the raw data into image suitable for our machine learning algorithms, we performed spatial downsampling and smoothing which averaged over the microscopic, cell-level variations within the data.
For the pair-rule-gene images, we performed a gaussian smoothing with $\sigma = 6 \mu$m.
For E-cadherin, which appears at cell junctions and therefore is more spatially inhomogeneous, we smoothed over a larger window of 
$\sigma = 15 \mu$m.
All fields were downsampled by a factor of 8.5, leaving $200\times 236$ images with spatial dimension $452\times 533 \mu$m.

In order to characterize junctional myosin, we first performed cytosolic normalization which defines a myosin concentration relative to the average level in the cytoplasm~\cite{gustafsonPatternedMechanicalFeedback2022,lefebvre_geometric_2023}.
The cytoplasm intensity $I_c$ is computed via a top-hat transform of the raw intensity $I$ over a single-cell region. The cytosolic-normalized signal is then
\begin{equation}
    I_n = \frac{I - I_c}{I_c}
\end{equation}
To characterize junctional anisotropy, we used the radon transform technique introduced in~\cite{streichanGlobalMorphogeneticFlow2018}. 
The radon transform method integrates the signal intensity within small windows along lines of varying orientation.
This maps bright cell edges to peaks in a Radon plane whose coordinates define a position and orientation for the edge. 
We use the maximum in the Radon plane to define the local orientation angle $\theta$ and director $\mathbf{n} = (\cos\theta, \ \sin \theta)$. As these orientation angles are defined modulo $180^{\circ}$, we define myosin anisotropy using a nematic tensor
\begin{equation}
    \mathbf{m} = m\, \mathbf{n} \mathbf{n}^T = 
    m \begin{pmatrix}
        \cos^2 \theta & \cos \theta \sin \theta \\
        \cos \theta \sin \theta & \sin^2 \theta
    \end{pmatrix}
\end{equation}
As with the E-cadherin data, we applied gaussian smoothing to the myosin tensor over a window $\sigma = 15 \mu$m and downsampled by a factor of 8.5. This produced a four-channel $236\times200$ image with spatial dimension $452\times 533 \mu$m, where each channel represented a component of the myosin tensor $\mathbf{m}$. 

\subsubsection{Ensemble time alignment}

\begin{figure}
    \centering
    \includegraphics[width=0.4\textwidth]{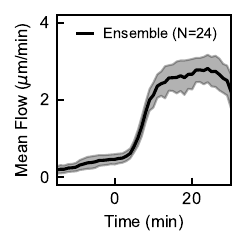}
    \caption{Plot of ensemble-averaged tissue flow magnitude over time following time-alignment of each embryo in the atlas.}
    \label{fig:ensemble_flow}
\end{figure}

To compare behavior across different embryos, we used time-alignment procedures reported previously for the \textit{Drosophila} atlas~\cite{mitchellMorphodynamicAtlasDrosophila2022}.
Briefly, all PRG datasets were co-stained with an antibody against Runt.  Each individual image could then be aligned by comparing the position and geometry of the Runt stripe pattern at each time point.
To align the live-imaged datasets of cytoskeletal proteins, we computed the spatially-averaged flow magnitude $\langle v \rangle$ and selected a constant offset for each embryo that minimized the difference between $\langle v \rangle(t)$ and that computed for a reference embryo. SI Fig.~\ref{fig:ensemble_flow} shows the average flow magnitude over time for the ensemble of aligned time series' for each embryo, demonstrating the reproducibility across embryos. 

\subsection{Deep Neural Networks}

\subsubsection{Deep Neural Network Architecture}

To forecast from initial conditions, we use an encoder-decoder architecture with a residual recurrent block at the latent bottleneck~\cite{colenMachineLearningActivenematic2021}. The network learns a set of latent variables from the inputs via a compressing convolutional encoder and computes the dynamics of those variables using a LSTM cell. Next the network transforms the latent variable trajectory into a spatially-resolved flow field via an upsampling convolutional decoder.  
At the latent bottleneck, we use the VAE reparameterization trick~\cite{Kingma.Welling2022} to enforce a smooth latent space. This means that for each latent parameter, the encoder computes a mean and variance of a latent probability distribution. During training, the latent vector is sampled from this distribution rather than computed deterministically from the inputs. All NNs were implemented using Pytorch~\cite{Paszke.etal2019}.

\begin{table}
    \centering
    \begin{tabular}{c|c c c| c}
         Module & Layer & Data Size & Channels & Details \\
         \hline
         \hline
         Encoder & Conv2d & $236\times200$ & $C\rightarrow32$ & k4$\times$4, stride 4 \\
         & $2\times$ConvNext & $59\times50$ & $32\rightarrow32$ & \\
         & Conv2d & & $32\rightarrow64$ & k$2\times2$, stride 2 \\
         & $2\times$ConvNext & $29\times25$ & $64\rightarrow64$ & \\
         & Conv2d & & $64\rightarrow128$ & k$2\times2$, stride 2 \\
         & $2\times$ConvNext & $14\times12$ & $128\rightarrow128$ & \\
         & Conv2d & & $128\rightarrow256$ & k$2\times2$, stride 2 \\         
         & $2\times$ConvNext & $7\times6$ & $256\rightarrow 256$ & \\
         \hline
         Latent & Linear & & $256\times7\times6 \rightarrow 2\times32$ & outputs $(\mu, \sigma^2)$\\
         & VAE trick & & & $\mathbf{z} = \mathcal{N}(\mu, \sigma^2)$ \\
         \hline
         Evolver & LSTM & & $32\rightarrow32$ & $\mathbf{z}_{t+1} = \mathbf{z} + LSTM(\mathbf{z})$ \\
         & & & & 2 layers, $h=64$ \\
         \hline
         Decoder & ConvTranspose2d & $7\times6$ & $32\rightarrow256$ & k$2\times2$, stride 2 \\
         & $2\times$ConvNext & $14\times12$ & $256\rightarrow 256$ & \\
         & ConvTranspose2d & & $256\rightarrow128$ & k$2\times2$, stride 2 \\
         & $2\times$ConvNext & $29\times25$ & $128\rightarrow 128$ & \\
         & ConvTranspose2d & & $128\rightarrow64$ & k$2\times2$, stride 2 \\
         & $2\times$ConvNext & $59\times50$ & $64\rightarrow 64$ & \\
         & ConvTranspose2d & & $64\rightarrow32$ & k$4\times4$, stride 4 \\
         & $2\times$ConvNext & $236\times200$ & $32\rightarrow 32$ & \\
         & Conv2d & & $32\rightarrow2$ & k$1\times1$
         
    \end{tabular}
    \caption{Forecasting NN architecture.}
    \label{tab:nnarch}
\end{table}

We use ConvNext blocks~\cite{Liu.etal2022} in our NN architecture. A standard convolutional layer learns $k\times k$ kernels for each input and output channel, meaning a $N\rightarrow N$ convolutional layer has $k^2 N^2$ tunable weights. In contrast, a ConvNext block learns spatial kernels that act only on each input channel via a grouped convolution, and then mixes each channel using $1 \times 1$ kernels. In the second step, it uses an inverse bottleneck structure with expansion factor $B$, which is equivalent to applying a fully-connected NN with hidden size $B\times N$ at each point in space. A ConvNext block uses $k^2 N + 2 B N^2$ weights. In our network, which uses $k=7$ and $B=4$, each block requires approximately 5 times fewer weights than an equivalent standard convolutional layer.  

\begin{table}
    \centering
    \begin{tabular}{c|c| c c c}
          Layer & Details & Channels & Kernel & Groups \\
         \hline
         \hline
         Conv2d & Mix space & $N \rightarrow N$ & 7x7 & N \\
         LayerNorm & & & \\
         Conv2d & Mix channels & $N \rightarrow 4N$ & 1x1 & 1 \\
         GELU & Nonlinear activation & & & \\
         Conv2d & Inverse bottleneck & $4N \rightarrow N$ & 1x1 & 1 \\
         Dropout & p=0.2 & & &
    \end{tabular}
    \caption{ConvNext Block architecture}
\end{table}

\subsubsection{Deep Neural Network Training}

We train for 100 epochs using the Adam optimizer with a learning rate of $\lambda=10^{-4}$. We use a learning rate scheduler which decreases the learning rate by a factor of 10 if the validation loss has not improved for 10 epochs. Each training sample is a randomly selected trajectory of length $2 + \mathcal{N}(0, 3)$ minutes. The inputs are a set of biological markers and the outputs are the flow field $\mathbf{v}$. Each embryo was imaged for only one biological marker, so we use ensemble averages in order to train networks to predict from multiple inputs. Here, we generate ensembles by aggregating all marked embryos with a timepoint within 1 minute of the live-imaged initial condition, and averaging over 2 randomly selected embryos in that ensemble.
During training, the network learns to minimize the following loss function
\begin{align}
    \mathcal{L} = \sum_t \text{Res}\big[\mathbf{v}_{NN}(t), \mathbf{v}_{Exp}(t) \big] + \beta D_{KL} \big[ \mathcal{N}(\mu, \sigma^2)  || \mathcal{N}(0, 1) \big]
    \label{eq:jointloss}
\end{align}
\begin{align}
    \text{Res}(\mathbf{u}, \mathbf{v}) = \frac{ \langle \mathbf{u}^2 \rangle \mathbf{v}^2 + \mathbf{u}^2 \langle \mathbf{v}^2 \rangle - 2 \mathbf{u} \cdot \mathbf{v}\sqrt{\langle \mathbf{u}^2 \rangle \langle \mathbf{v}^2 \rangle } }{2 \langle \mathbf{u}^2 \rangle \langle \mathbf{v}^2 \rangle}
    \label{eq:residual}
\end{align}
This joint loss function contains two terms. The first is a reconstruction loss based on the residual metric defined in~\cite{streichanGlobalMorphogeneticFlow2018}. The second term is a Kullback-Leibler divergence between the latent variables and a normal distribution. 
The coefficient $\beta$ determines the relative importance during training between the KL divergence term and the reconstruction residual. When training the models in Fig.~\ref{fig:deepNN}, we set $\beta = 0$. When re-training the models for closed-loop forecasting (Fig.~\ref{fig:sindy}), we set $\beta = 10^{-4}$ to better condition the latent space for handling artifacts that might occur during the numerical procedure. 

\subsection{Identifying predictive proteins using neural networks}
\label{sec:predictive_proteins}

\begin{figure}[ht]
    \centering
    \includegraphics{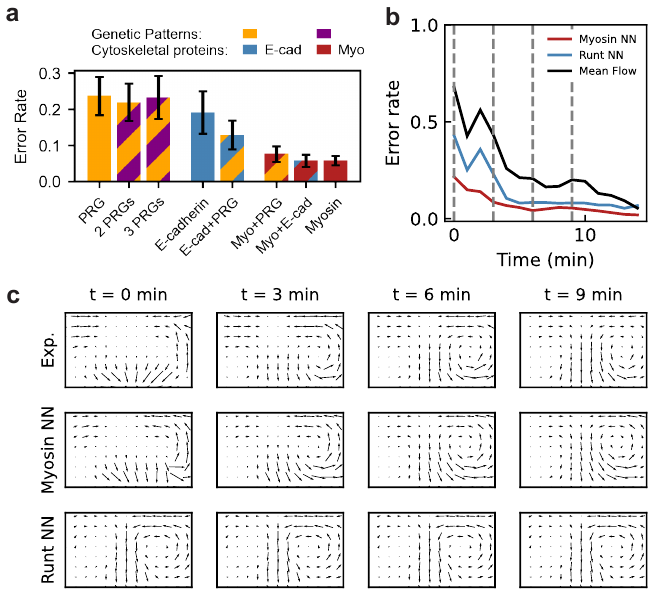}
    \caption{\textbf{Myosin is an optimal predictor of tissue flow.}
        (\textbf{a}) Average test error rate for NNs trained on PRGS, cytoskeletal proteins, and combinations. Networks trained using myosin as input outperformed all others.  
        (\textbf{b}) Error-rate for single predicted trajectories for networks trained on myosin (red) and a PRG (runt, blue). Error rates were calculated against the ensemble-averaged flow trajectory. We also plot the error-rate for a "mean-field" prediction (black), which is given by the time- and embryo-averaged flow fields over the entire dataset.
        (\textbf{c}) Snapshots of experimental and predicted flow fields at three-minute intervals marked by grey lines in (\textbf{b}). While the myosin-trained network predicts a transition from a VF flow pattern to a vortical GBE configuration, the PRG-trained network predicts a constant flow field over time.
    }
    \label{fig:protein_comparison}
\end{figure}

Using a morphodynamic atlas~\cite{mitchellMorphodynamicAtlasDrosophila2022}, we trained NNs to predict tissue flow from PRGs and cytoskeletal proteins. Myosin yielded the best predictions (6\% error on unseen embryos), while E-cadherin (19\%) and the live-imaged PRGs Runt (24\%) or Even-Skipped (23\%) performed worse. To train on multiple patterns, we supplemented each movie with an ensemble average of a second time-aligned field. Training on more than one PRG (including the static-imaged Fushi-Tarazu, Paired, and Sloppy-Paired) yielded no improvement, while supplementing E-cadherin with myosin or a PRG did improve results. Including other fields alongside myosin produced no improvement beyond the accuracy achieved by myosin alone (see Fig.~\ref{fig:protein_comparison}a).

To contextualize the error rates reported in Fig.~\ref{fig:protein_comparison}a, we used myosin-trained and PRG-trained NNs to predict 15-minute trajectories from initial conditions coinciding with VF onset ($t = 0$). In Fig.~\ref{fig:protein_comparison}b-c, we plot the error rates over time and include sample snapshots of the predicted flow from each trajectory. The myosin-trained NN predicts a transition from a VF to a vortical GBE flow configuration. The PRG-trained NN gives a nearly-constant field which only slightly outperforms a constant "mean-field" prediction of the time- and ensemble-averaged flow field. Both achieve low error rates after the onset of GBE where the flow is quasi-static but have higher error rates during VF and GBE onset when the flow changes rapidly. Indeed, the network which predicts using myosin obtains a lower error rate over the entire trajectory (7\%) than either the PRG (14\%) or "mean-field" (26\%) predictions.

\subsubsection{Biological perspective} 

While myosin may be all a NN needs to generate flow, the same is not true for the embryo. Myosin binds to actin and produces stresses which propagate through the tissue across cell junctions. The NN results do not imply that other cytoskeletal proteins are unnecessary for axis elongation. Rather, the NN learned to use myosin as a proxy for this complex microscopic machinery, and can infer the necessary information to predict tissue-scale flow from myosin alone.

\subsubsection{Mechanical perspective} 
The NNs reaffirm the pivotal role of myosin in driving flow during axis elongation but they are not constrained by physics.
The features they extract from myosin can be more complex than what might enter a hydrodynamic model via a local gradient expansion.
Indeed, a recent study found that distilling a NN which predicted cell mechanics from proteins into a continuum mechanical model required adding new information -- displacements and cell geometry -- which the NN never saw~\cite{Schmitt.etal2023}.
Similarly, a hydrodynamic model for \emph{Drosophila} might require more than just myosin. 

\subsection{Principal component analysis}

To perform the decomposition into sparse components, we constructed a pipeline using the scikit-learn Python library~\cite{Pedregosa.etal2018}. The pipeline accepted the processed image data and iteratively fit and transformed using the following steps

\begin{enumerate}
    \item StandardShaper - a custom class to transform input data to size $[N_{samples}, C, H, W]$, where $H, W$ are the image dimensions, $C = 4$ for tensor data, $C=2$ for flow fields, and $C=1$ for scalar fields. 
    \item LeftRightSymmetrize - a custom class which symmetrizes the image along the lateral direction, to prevent any left-right asymmetry from dominating the principal components
    \item Masker - a custom class which applies a specified crop and mask to the data. In all cases we cropped 10 pixels (22 $\mu$m) from the boundary in each direction to eliminate distortion at anterior/posterior poles, as well as behavior at the VF. We also masked out regions by the cephallic furrow and posterior midguts at later timepoints.
    \item ModifiedStandardScaler - Unlike traditional standardization which centers the data by subtracting the mean, we subtract the ensemble average of all timepoints before VF invagination. Note that by not centering the data to have zero mean before decomposing, this pipeline is not \textit{technically} PCA.
    \item TruncatedSVD - Singular Value Decomposition truncated to $N=16$ components.
\end{enumerate}

\subsection{SINDy fitting procedure and library construction}
\label{sec:sindy_details}

\subsubsection{Library construction}

For each field in the library ($\mathbf{m},\, \mathbf{v},\, c$), we first projected onto the first $N$ PCA components,  where $N = $ the number of components that explain 95\% of variance. 
From these smoothed projections of the raw data, we computed the following libraries of scalar and tensor terms:
\begin{align}
    \mathcal{L}_{c} &= \bigg\{ 1,\ c,\ \text{Tr}(\mathbf{m}),\ v^2, \nabla \cdot \mathbf{v} \bigg\} \\
    \mathcal{L}_{m} &= \bigg\{ \mathbf{m},\  \mathbf{m} \text{Tr}(\mathbf{m}),\ \bm{\Omega},\  \mathbf{E}, \ \bm{\Gamma}^{\text{DV}} \bigg\}
\end{align}
Here $\bm{\Omega} = (\nabla \mathbf{v} - \nabla \mathbf{v}^T) / 2$ is the vorticity tensor, and $\mathbf{E} = (\nabla \mathbf{v} + \nabla \mathbf{v}^T) / 2$ is the strain rate tensor. We also included a static DV-aligned source $\bm{\Gamma}^{\text{DV}}$ in the tensor library to enable a tissue-scale geometric stress as in~\cite{lefebvre_geometric_2023}. We used these base sets to generate composite libraries using the following operations
\begin{enumerate}
    \item Advective terms:  $(\mathbf{v} \cdot \nabla) c$, $(\mathbf{v} \cdot \nabla) \mathbf{m}$
    \item Symmetric couplings of tensors in $\mathcal{L}_m$. Note that this creates the co-rotation term $[\bm{\Omega}, \mathbf{m}]$
    \item Nonlinear couplings of scalars in $\mathcal{L}_c$
    \item Outer product of tensor library and $\{ 1, c \}$ to include E-cadherin-modulated terms
    \item Removal of any terms higher than first order in derivatives
\end{enumerate}
The second operation enables myosin-myosin and myosin-strain feedback while the fourth operation enables cadherin-modulation of myosin dynamics. 
We used these operations to generate a candidate library and stored it in an H5F filesystem for faster access during repeated training operations. We also stored the projections of each field onto the principal components. Finally, we included the time derivatives of the projected fields, computed using smoothed finite differences with a window size of 7 minutes, for use in the SINDy fitting procedure~\cite{bruntonDiscoveringGoverningEquations2016}.

In the atlas, myosin and E-cadherin were measured on different embryos. To compute terms which couple $\mathbf{m}$ and $c$, we leveraged time-aligned ensemble averages enabled by the \textit{Drosophila} atlas.
Specifically, to compute a term like $c \, \mathbf{m} (t=t_0)$ for an embryo live-imaged for myosin, we used $c = \langle c(|t-t_0| < 1) \rangle$. A similar replacement was used for terms like $c \, \text{Tr}(\mathbf{m})$ for E-cadherin-imaged embryos.

\subsubsection{SINDy fitting}

To fit equations to the data using the above libraries, we collected each term and standardized them so that they had zero mean and unit variance over the entire dataset. 
We used the SINDy procedure and the Sequentially-Thresholded Least Squares (STLSQ) selection algorithm~\cite{bruntonDiscoveringGoverningEquations2016}. Rather than fit equations to the bare time derivative, we instead fit to the material derivative by adding the relevant terms to the left hand side before optimization. That is, we fit a myosin equation to the target $\dot{\mathbf{m}} + (\mathbf{v} \cdot \nabla)\mathbf{m} + [\bm{\Omega}, \mathbf{m}]$, with the latter two terms removed from the candidate library. Similarly, we fit a E-cadherin equation to the target $\dot{c} + (\mathbf{v} \cdot \nabla)c$ and removed the latter term from the candidate library.

STLSQ iteratively performs a regularized least squares minimization procedure. At each iteration, it removes any library terms whose standardized coefficients are below a thresholding parameter $\tau$, which we set to $\tau = 0.01$. 
This process is repeated until the result converges. 
For the least squares fitting, we used L2 (ridge) regularization with a ridge parameter $\alpha = 10$. 

The SINDy procedure is not deterministic, especially when the data is noisy and the library is large or contains several nearly equivalent terms. 
To account for this, we repeated the fitting procedure and kept terms that were present in an ensemble average over these runs. 
The terms present in (\ref{eq:myosin_sindy}) were found via an ensemble average over 10 trials, each trained on 10\% of the dataset.

When performing an ensemble average of trials on a large library, the magnitudes of each coefficient may be suppressed by the presence of subleading terms that appeared stochastically in different models within the overall ensemble.
For this reason, we repeated the fitting procedure using a library restricted to only the terms present in (\ref{eq:myosin_sindy}). 
We used the resulting equation, listed below, to integrate the model and generate the results in Fig.~\ref{fig:sindy}.
\begin{multline}
        \partial_t \mathbf{m} + (\mathbf{v} \cdot \nabla) \mathbf{m} + [\Omega, \mathbf{m}] = 
        -0.06 (1 - 0.9\, c) \ \mathbf{m} + 
        0.56  (1 - 0.7\, c) \ m\, \mathbf{m} + \\
        0.49  (1 + 0.6\, c) \ E\, \mathbf{m} + 
        0.05  (1 - 0.8\, c) \ m\, \bm{\Gamma}^{\text{DV}}
        \label{eq:integratedmyosin}
\end{multline}

\subsubsection{Hybrid integration of cytoskeletal and tissue dynamics}

To integrate the cytoskeletal dynamics, we used a hybrid model which forecasts myosin and E-cadherin using (\ref{eq:cadherin_sindy})-(\ref{eq:myosin_sindy}) and predicts instantaneous tissue flow using a NN (Fig.~\ref{fig:sindy}a).
At each time step, we computed the field $\mathbf{v}$ from the current myosin field $\mathbf{m}$ using a myosin-trained NN with the same architecture as in Fig.~\ref{fig:deepNN}e-h. 
To better handle artifacts that may arise through the numerical integration, we re-trained the network for 500 epochs with a higher weight $\beta=10^{-4}$ on the KL-divergence term in (\ref{eq:jointloss}). 
In order to translate the instantaneous flow rather than forecast, we omitted the recurrent LSTM layer (Evolver module in Table~\ref{tab:nnarch}) and only applied the Encoder, Latent, and Decoder modules to the $\mathbf{m}$ field.
Because the network was trained using the normalized residual presented in~\cite{streichanGlobalMorphogeneticFlow2018}, it predicts flow fields subject to an overall scale factor. Prior to integration, we calibrate this scale factor once by applying the NN to the experimental myosin field and comparing its output magnitudes to those measured in experiment. The NN is then applied to the myosin field predicted by (\ref{eq:myosin_sindy}) at each time step. 
After the NN predicts the instantaneous flow, we computed the derivatives $\dot{\mathbf{m}},\ \dot{c}$ according to the machine-learned dynamical equations.
To eliminate artifacts, we performed two post-processing steps.
First, we zeroed each time derivative at the anterior and posterior poles (15 pixels or 33 $\mu$m from each pole) in order to eliminate artifacts due to distortion that may be amplified further when taking spatial gradients, following~\cite{streichanGlobalMorphogeneticFlow2018}.
Second, we applied a gaussian filter ($\sigma = 5$ pixels $= 11\mu$m) to each field in order to smooth over upsampling artifacts introduced by the NN. 
We performed a fourth order Runge-Kutta integration using the torchdiffeq library~\cite{Chen.etal2019} for 30 minutes, using an initial condition of the ensemble-averaged myosin field 10 minutes prior to VF and ending approximately 14 minutes after GBE onset. Following~\cite{streichanGlobalMorphogeneticFlow2018}, we compare predictions using the normalized error metric (\ref{eq:residual}).

While past work has proposed that a Stokes equation can determine tissue flow from the global myosin distribution~\cite{streichanGlobalMorphogeneticFlow2018}, we have instead used a NN to predict instantaneous flow. The reason for this is that~\cite{streichanGlobalMorphogeneticFlow2018} required adding a ventral cut to the 3D embryo mesh to model cell internalization during VF. This spontaneous change in geometry is a distinct morphogenetic event that is beyond the ability of our hydrodynamic models to predict. Instead, we used a NN which learned to predict flow observed in experiment from myosin directly. It does not need to know the system's changing geometry, but instead infers its effects from features of the myosin pattern. Using a NN kept the dynamical system closed and autonomous such that it could be integrated from initial conditions.

\subsection{Minimal active matter modeling}
\label{sec:modeling}

\subsubsection{Myosin alone cannot reproduce GBE dynamics}

Using minimal continuum models, we aim to reproduce (i) a long-lived vortex flow pattern and (ii) a long-lived myosin DV gradient (see Fig.~\ref{fig:PCA}c'-f'). As a starting point, we consider the following equations.
    \begin{equation}
        \mu \nabla^2 \mathbf{v} - \nabla P + \alpha \nabla \cdot  \mathbf{m} = 0
        \label{eq:stokes}
    \end{equation}
    \begin{equation}
        D_t \mathbf{m} = A\  \mathbf{m} + B\  \mathbf{m}\, \text{Tr}(m) + D\ \nabla^2 \mathbf{m}
        \label{eq:myodiffuse}
    \end{equation}

Eq.~\ref{eq:stokes} is an incompressible ($\nabla \cdot \mathbf{v} = 0$) Stokes flow driven by myosin active stress $\nabla \cdot \textbf{m}$ \cite{streichanGlobalMorphogeneticFlow2018}. To account for myosin anisotropy on cell-cell junctions, we use the local orientation angle $\theta(\mathbf{r})$ to write a director $\mathbf{n} = (\cos \theta, \sin \theta)$ and analogue to a nematic order parameter $\mathbf{m} = m\, \mathbf{n} \mathbf{n}^T$ ($m$ is the local myosin intensity).
Eq.~\ref{eq:myodiffuse} describes myosin dynamics in terms of $\mathbf{m}$ where the left hand side is the co-rotational derivative 
$D_t \mathbf{m} = \partial_t \mathbf{m} + (\mathbf{v} \cdot \nabla) \mathbf{m} + (\bm{\Omega} \cdot \mathbf{m} - \mathbf{m} \cdot \bm{\Omega})$. 
The right hand side contains minimal terms modeling myosin feedback (albeit somewhat non-biological, as myosin is bound in cells and cannot directly diffuse).
This system of equations has two spatially-uniform fixed points $\mathbf{m} (\mathbf{r}) = \mathbf{m}_0$, with $\text{Tr}(m_0) = 0, -A/B$ (with constant myosin, $\mathbf{v} \rightarrow 0$ and $D_t \rightarrow \partial_t$), and the latter exists only when $AB < 0$. Unless $A,B > 0$, the system will evolve to a constant stable fixed point and produce no flow. Otherwise, the system is unstable to long-wavelength perturbations with wavenumber $q^2 < A / D$. 
In this case, the model produces a plausible myosin and flow pattern, but due to the instability it will grow without bound. In our simulation, the myosin anisotropy direction is inherited from a polarized, nearly homogeneous, experimental initial condition. 

\begin{figure}
    \centering
    \includegraphics[width=0.7\textwidth]{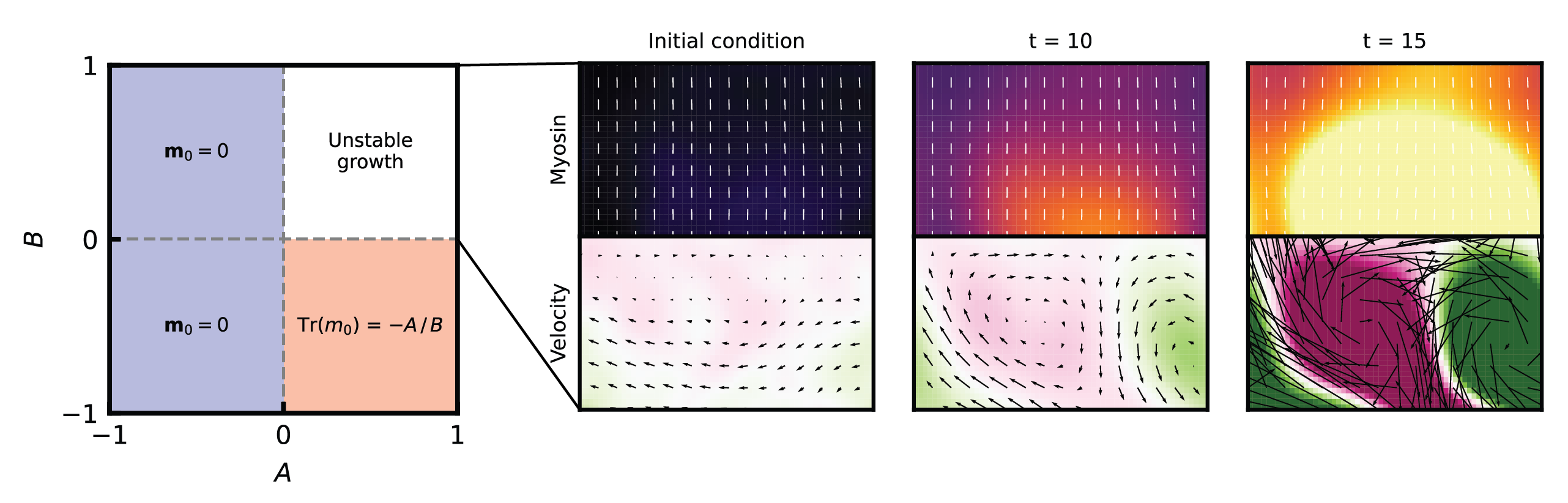}    
    \caption{\textbf{Myosin alone cannot reproduce GBE dynamics.}
        \textit{(Left)} Phase diagram for the system of equations Eq.~\ref{eq:stokes}-\ref{eq:myodiffuse}.
        \textit{(Right)} Prediction of pattern-forming in a myosin-only dynamical system enabled by a hydrodynamic instability. 
    }
    \label{fig:MyosinOnly}
\end{figure}

\subsubsection{Hydrodynamics with a control field}

We now add a stationary spatially-patterned field $c(\mathbf{r})$ which cross-regulates the myosin. 
\begin{equation} 
    \dot{m} + (\mathbf{v} \cdot \nabla) m = A\left[ 1 - k\, c(\mathbf{r})\right] m + B m^2
    \label{eq:myocad_conc}
\end{equation}
While myosin includes concentration and orientation dynamics, we focus on the former in Eq.~\ref{eq:myocad_conc}. A non-uniform myosin pattern $m$ can be maintained by $c(\mathbf{r}) = (A\, m + B\, m^2 - \mathbf{v} \cdot \nabla m) / k \,m$. This stationary configuration replaces the non-trivial fixed point (red) in the above phase diagram. As an example, we numerically computed a pattern $c^{\star}$ which would maintain the GBE configuration at $t = 20$ minutes post-VF and found it is DV-graded similar to E-cadherin. We integrated the system using the \textit{experimental} E-cadherin field, and observed it evolve to a patterned myosin 
field. 

Orientational dynamics ---myosin rotation by vorticity--- prevents this pattern from being a true fixed point. Over longer timescales, the pattern disappears after GBE flow destroys the orientational order of myosin. In the subsequent sections, we show mechanical feedback slows this effect, enabling a longer-lived GBE state.

\begin{figure}
    \centering
    \includegraphics[width=0.7\textwidth]{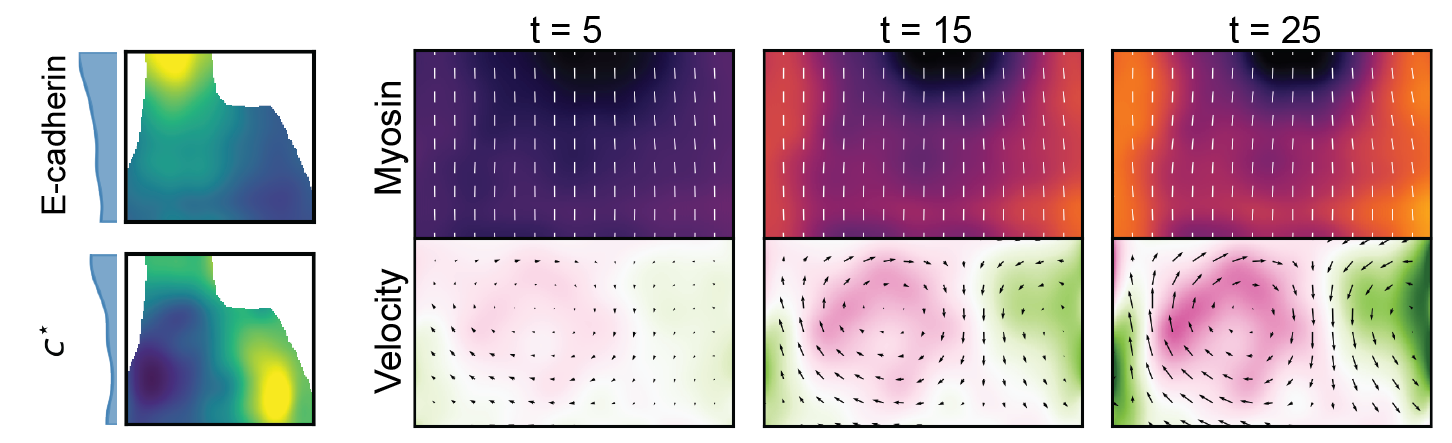}
    \caption{\textbf{Hydrodynamics with a control field.}
        \textit{(Left)} Comparison of an experimental E-cadherin field with a field $c^{\star}$ obtained by setting the RHS of Eq.~\ref{eq:myocad_conc} to zero. Both fields are DV-graded with higher levels dorsally. 
        \textit{(Right)} Integration of Eq.~\ref{eq:myocad_conc} with the experimental E-cadherin field produces a myosin pattern which is longer lived than one generated via a hydrodynamic instability (Fig.~\ref{fig:MyosinOnly}).
    }
    \label{fig:MyoCad}
\end{figure}

\subsubsection{Myosin anisotropy modifies hydrodynamics}

After focusing on the myosin concentration in previous sections, we now consider orientation dynamics. 
\begin{equation}
        \dot{\mathbf{m}} + (\mathbf{v} \cdot \nabla) \mathbf{m} + (\bm{\Omega} \cdot \mathbf{m} - \mathbf{m} \cdot \bm{\Omega}) = 
        A \left[ 1 - k c(\mathbf{r}) \right]\, \mathbf{m} + B\, \mathbf{m} \text{Tr}(m)
        \label{eq:myosin_tensor_dyn}
\end{equation}
 Eq.~\ref{eq:myosin_tensor_dyn} has a co-rotation term which describes how myosin orientation deflects due to gradients in flow. Note that we have dropped the unbiological diffusion term $D\nabla^2\mathbf{m}$. We separate concentration and orientation dynamics using the stream function $\psi\ (\mathbf{v} = \nabla \times \psi)$ and orientation matrix $\mathbf{D}_{\theta} = \mathbf{n} \mathbf{n}^T$. 
\begin{equation}
        \underbrace{\bigg[ \dot{m} - \nabla \psi \times \nabla m \bigg]}_{\text{Concentration dynamics}} \mathbf{D}_{\theta} +
        m 
        \underbrace{\bigg[ \dot{\theta} - \nabla \psi \times \nabla \theta + \frac{1}{2} \nabla^2 \psi \bigg]}_{\text{Orientation dynamics}} \nabla_{\theta} \mathbf{D}_{\theta} = \left[ A (1 - k c) m + B m^2 \right] \mathbf{D}_{\theta}
        \label{eq:decomposed_dyn}
\end{equation}
This decomposition shows that the RHS creates a stationary concentration pattern via a fixed source $c(\mathbf{r})$, but the orientation will always vary, as argued in Ref.~\cite{lefebvre_geometric_2023}. Simulations support this result. We integrate from an initial condition with a simple control field profile $c^{\star} = c_0 + c_1 \cos (\frac{2\pi x}{L_x} + \frac{2\pi y}{L_y})$ and myosin $m = m_0,\ \theta = \pi/2$. The myosin concentration approaches the fixed point $m^{\star}$ defined by the control field such that
the RHS in Eqs.~\ref{eq:myosin_tensor_dyn}-\ref{eq:decomposed_dyn} $= 0$ but this GBE-like configuration is short-lived. Vorticity creates sharp gradients in the myosin orientation, producing unrealistic and rapidly-diverging flow patterns. 

\begin{figure}
    \centering
    \includegraphics[width=0.7\textwidth]{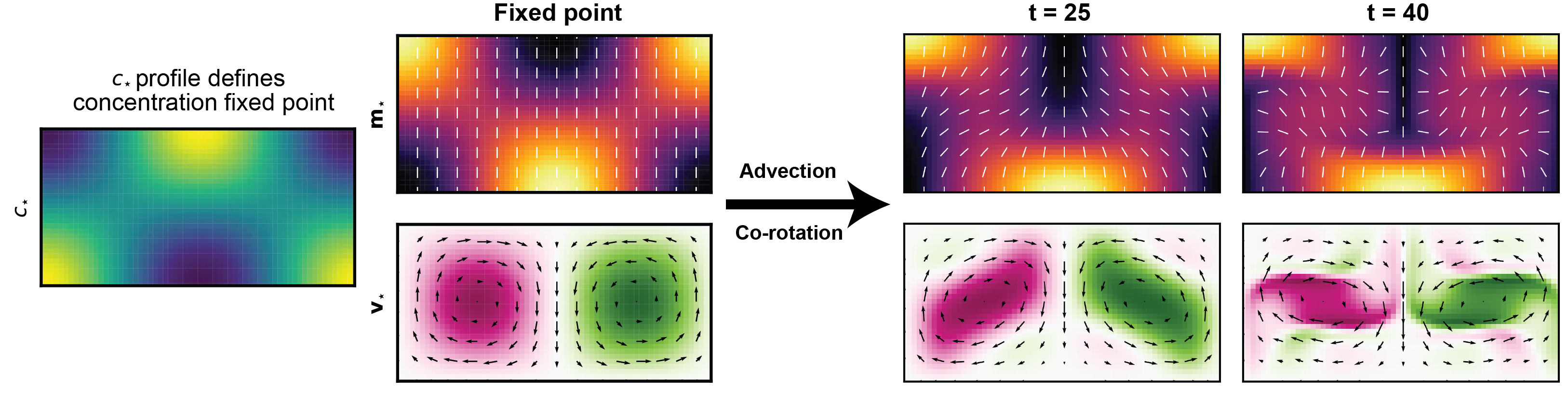}
    \caption{\textbf{Orientational dynamics destroy fixed point over time.}
        \textit{(Left)} For a simplified control field profile $c^{\star}$, the system evolves to a quasi-fixed point $m^{\star}$ which qualitatively reproduces GBE-like flow. 
        \textit{(Right)} The nonlinear advection and co-rotation terms in Eq.~\ref{eq:decomposed_dyn} create sharp gradients in myosin orientation, eventually destroying the GBE flow pattern. 
    }
    \label{fig:orientation_dynamics}
\end{figure}

\subsubsection{Mechanical feedback slows destruction of orientational order}

The flow caused by myosin rotates junctional orientations and ultimately destabilizes the GBE configuration in Eq.~\ref{eq:myosin_tensor_dyn}. However, experiments have shown that tissue flow is nearly static during GBE~\cite{mitchellMorphodynamicAtlasDrosophila2022}, suggesting that additional hydrodynamic terms may
be necessary to agree with biology. Recent work has demonstrated that mechanical feedback can recruit myosin to cell-cell junctions~\cite{gustafsonPatternedMechanicalFeedback2022}. Including a strain-coupling term proportional to $\{\mathbf{m} , \mathbf{E}\} = \mathbf{m} \cdot \mathbf{E} + \mathbf{E} \cdot \mathbf{m}$ appears to slow the rate at which $\theta$ changes and preserves the GBE flow pattern nearly twice as long. To understand this, consider dynamics about $m = m^{\star},\ \theta = \pi / 2$. Here, the new term becomes $\{ \mathbf{m}, \mathbf{E} \} = - m \mathbf{D}_{\theta} \nabla^2_{xy} \psi$, and the dynamics are given by Eq.~\ref{eq:mech_feed}. 
\begin{equation}
    \bigg[ \dot{m} - \nabla \psi \times \nabla m + k_E \,m \nabla^2_{xy} \psi \bigg] \mathbf{D}_{\theta} + 
    m \bigg[ \dot{\theta} + \frac{1}{2} \nabla^2 \psi \bigg] \nabla_{\theta} \mathbf{D}_{\theta} = 0
    \label{eq:mech_feed}
\end{equation}
At first, it is surprising that $\{ \mathbf{m}, \mathbf{E} \}$ preserves orientational order, as it only suppresses $\dot{m}$ rather than $\dot{\theta}$. 
Consider instead $\ddot{\theta} = -\frac{1}{2} \nabla^2 \dot{\psi}$, the rate at which $\dot{\theta}$ grows. Differentiating the Stokes equation yields
\begin{equation}
    \nabla^4 \dot{\psi} = -2 \nabla^2 \ddot{\theta} = \frac{\alpha}{\mu} \left[\nabla^2_{xy} \dot{m}  -(\nabla^2_x - \nabla^2_y) m \dot{\theta}\right]
\end{equation}
In this Poisson equation for $\ddot{\theta}$, mechanical feedback suppresses the source term containing $\dot{m}$.
Rather than eliminate $\dot{\theta}$, mechanical feedback adjusts the concentration (and flow) dynamics to slow $\dot{\theta}$'s growth. This prevents \textit{accelerating} destruction of orientational order (see plot of correlation length $\ell_\theta$ over time) and allows the system to maintain GBE flow longer. In the subsequent section, we interpret the term $\{\mathbf{E}, \mathbf{m}\}$ in more detail. 

\begin{figure}
    \centering
    \includegraphics[width=0.4\textwidth]{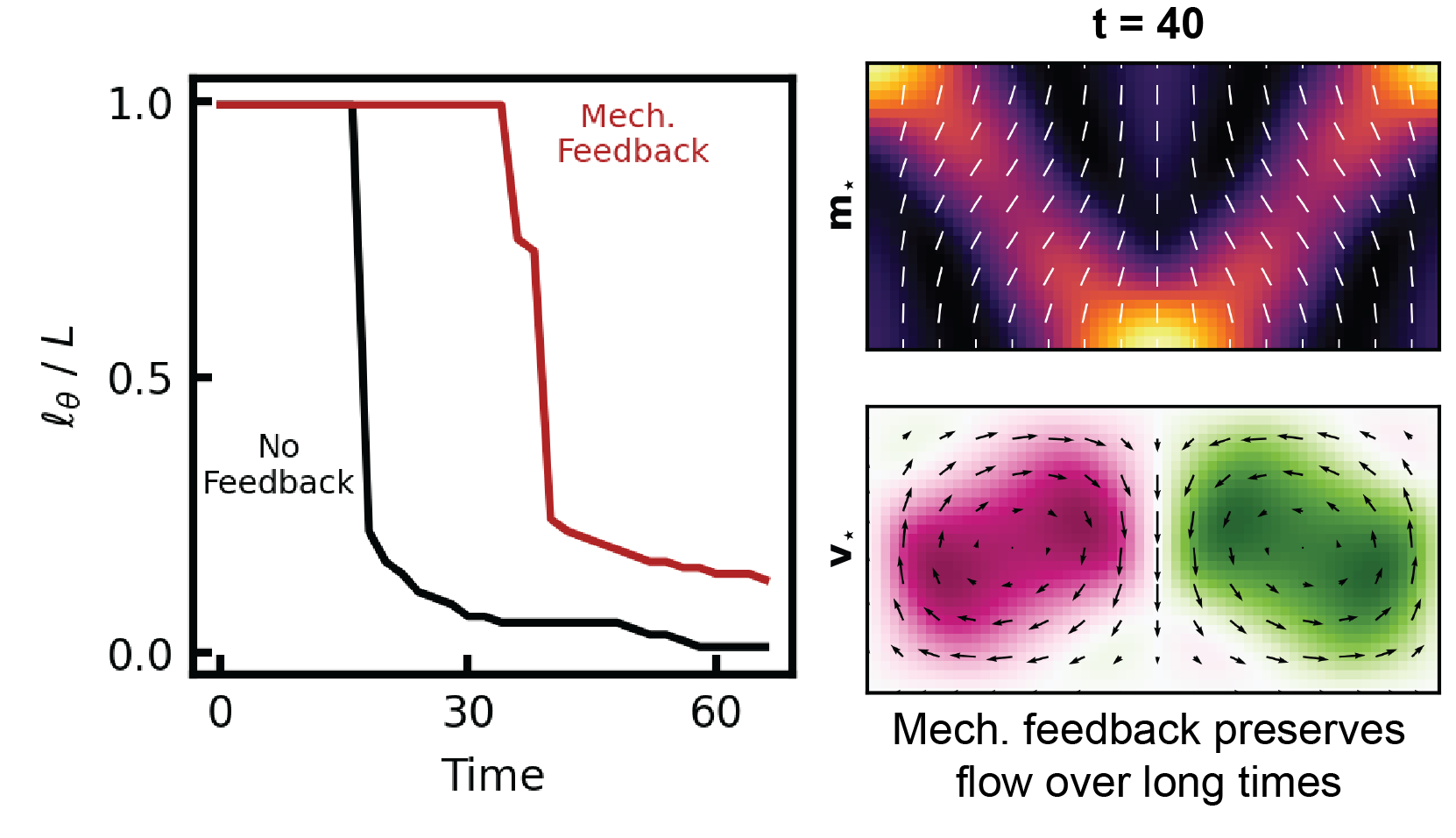}
    \caption{\textbf{Mechanical feedback term preserves orientational order.}
        \textit{(Left)} Comparison of correlation lengths over time in simulations with (red) and without (black) mechanical feedback terms. The mechanical feedback preserves orientational ordering for nearly twice as long.
        \textit{(Right)} Plot of simulated myosin and flow fields at $t=40$ (A.U.) in a model with the mechanical feedback term $\{ \mathbf{m}, \mathbf{E} \}$. Compared to Fig.~\ref{fig:orientation_dynamics}, the modified model better preserves the vortical flow structure associated with germband-extension.
    }
    \label{fig:mech_feedback}
\end{figure}

\subsection{Interpretation of of machine-learned continuum equations}
\label{sec:interpretation}

Here we connect the learned continuum equations to previous models of myosin dynamics during GBE. The right hand sides of the SINDy equations Eqs.(\ref{eq:cadherin_sindy})-(\ref{eq:myosin_sindy}), are the usual convective derivatives, because myosin $\mathbf{m}$ and E-cadherin $c$ are localized to cells that are transported by tissue flow velocity $\mathbf{v}$. As explained in the preceding section, the myosin pattern is strongly anisotropic, which is why we describe it using a nematic tensor, while E-cadherin, appears localized to cell-cell interfaces in an isotropic manner and can be described by a scalar density.

We note that the SINDy equations combine several terms, and are non-linear, which makes a complete analysis, in particular of the instabilities, challenging. 
Distinguishing strain-rate and stress (tension) effects, for example, is difficult as they are typically correlated. 
Moreover, SINDy solves an underdetermined regression problem with potentially many equivalent solutions, meaning interpreting individual terms in (\ref{eq:myosin_sindy}) must be done with caution.
Our equation is not intended to model other, distinct morphogenetic events that occur simultaneously with axis elongation, such as ventral furrow and posterior midgut invagination.
We do not attempt a complete analysis here and instead comment on the interpretation of the different terms in (\ref{eq:myosin_sindy}) and their connection to the prior literature.

\subsubsection{E-cadherin dynamics}

The right-hand side of the learned E-cadherin equation (\ref{eq:cadherin_sindy}) is 0. This means E-cadherin is conserved. Indeed, biologically, the turnover of E-cadherin, which is a trans-membrane molecule, is very slow \cite{Chen.etal2017}, and the E-cadherin pattern can be considered as conserved over a timescale of 20 minutes.

\subsubsection{Myosin dynamics: turnover}

We now turn to the myosin dynamics in equation (\ref{eq:myosin_sindy}). The term $-k_1 \mathbf{m}$ corresponds to constant detachment of myosin from junctions. Such detachment is indeed measured experimentally using FRAP \cite{lefebvre_geometric_2023}.

\subsubsection{Myosin dynamics: tension feedback}

The learned equations (\ref{eq:cadherin_sindy})-(\ref{eq:myosin_sindy}) contain additional RHS terms that model tension feedback, or recruitment of myosin on edges under higher tension. This mechanism was previously analyzed in Ref. \cite{lefebvre_geometric_2023}, which considered only the local orientation $\theta \in [0, \pi]$ of the myosin nematic tensor $\mathbf{m}$, and not its magnitude. It was shown that the behavior of $\theta(t, \mathbf{r})$ can be understood as a combination of advection (i.e. tissue rotation), detachment, and recruitment by a static source. These mechanisms lead to following equation for the full myosin tensor:
\begin{align}
    \partial_t \mathbf{m} + (v_k\nabla_k) \mathbf{m} + [\bm{\Omega}, \mathbf{m}] = -k_1 \mathbf{m} + k \cdot (\mathbf{m} + \bm{\Gamma}^\text{DV})
    \label{eq:lefebvre}
\end{align}
of which Eqs. (\ref{eq:cadherin_sindy})-(\ref{eq:myosin_sindy}) are an extension. 
The static source $\bm{\Gamma}^\text{DV}$ was hypothesized to be the surface stress that balances the internal turgor pressure in the embryo. Because the embryo is elongated, this stress is anisotropic, and because the embryo geometry does not change during tissue flow, it is static. 

Ref. \cite{brauns_epithelial_2023} analyzed GBE on a single-cell level and found that the dynamics of active tension $T$ on single edges (computed using tension inference), was well described by excitable tension recruitment:
\begin{align}
    \partial_t T = k T^n, \quad n >1
    \label{eq:brauns}
\end{align}
Tension feedback has previously been found experimentally \cite{Fernandez-Gonzalez2009,Yu2016}. The mesoscale tension recruitment term $k_2\, m\, \mathbf{m}$ can be viewed as a coarse-graining of Eq. \ref{eq:brauns}.

\subsubsection{Myosin dynamics: Strain rate feedback}

Ref. \cite{gustafsonPatternedMechanicalFeedback2022} has shown that junctions recruit myosin in response to deformation strain rate. The paper analyzes single junctions of length $\ell$ and myosin level $m$. Strain rate recruitment is modelled by :
\begin{align}
    \frac{\dot{m}}{m} = \alpha(\mathbf{r}) \frac{\dot{\ell}}{\ell}
    \label{eq:gustafson}
\end{align}
This type of feedback was first postulated theoretically as a way for cells to ensure convergence to mechanical equilibrium while maintaining tissue plasticity \cite{nollActiveTensionNetwork2017}.
The feedback coefficient $\alpha$ does not depend on the junction orientation, but it does depend on the position along the DV axis. It was found experimentally that $\alpha$ was low dorsally and high ventrally. To interpret this strain-rate feedback in the context of Eqs. \ref{eq:cadherin_sindy}-\ref{eq:myosin_sindy}, we first coarse-grain Eq.~\ref{eq:gustafson} to determine how it might appear in a continuum model. Next, we examine a discrepancy between the feedback coefficient in Eq.~\ref{eq:myosin_sindy} and the patterned mechanical feedback reported in~\cite{gustafsonPatternedMechanicalFeedback2022}. \\

\subsubsection{Coarse-graining single-cell level strain rate feedback}
We can connect the continuum equation terms involving the strain rate with this edge-level model.
The anticommutator $\{E, m\}_{ij} = (E_{ik} m_{kj} + m_{ik} E_{kj} )/2$ can be seen as an (inexact) coarse graining of Eq. \ref{eq:gustafson}. To see this, define the single-edge myosin tensor $m_{ij}^{e}$, computed from the junction orientation $\mathbf{n}$ and the myosin level $m$ of a junction $e$: $m^e_{ij} = m n_i n_j$. In presence of local strain rate $E$, the junction will become stretched depending on its relative orientation to the strain axes:
\begin{align}
    \frac{\dot{\ell}}{\ell} = \alpha n_i E_{ij} n_j =  \alpha \frac{m^e_{ij} E_{ji}}{m}
\end{align}
Second, if the orientation $\mathbf{n}$ is not perfectly aligned with the eigenvectors of $E$, the junction orientation will also be affected.
\begin{align}
    \partial_t \mathbf{n}_i = E_{ij} n_j -  (n_j  E_{jk} n_k ) n_i
\end{align}
The second term here ensures that $\mathbf{n}$ remains a unit vector. 
Because of its non-linear nature, this term leads to subtleties when coarse graining. Ignoring these issues, however, we can combine the two preceding equations into tensorial form:
\begin{align}
    \partial_t m^e_{ij} = \alpha \{E, m^e\}_{ij}
    \label{eq:continuum_strain_rate}
\end{align}
Another way to derive Eq. \ref{eq:continuum_strain_rate} is to model the local angular distribution $m(\theta)$ of myosin intensity as a function of edge orientation. This distribution gets shifted by the strain rate $E$, and the resulting equations can be expanded in a Fourier basis in $\theta$ and repackaged into a tensor. However, similar subtleties make this coarse graining inexact.

Eq. \ref{eq:continuum_strain_rate} is the term that is considered in Eq.~\ref{eq:mech_feed}, and the strain rate term $\mathrm{Tr}[E]\mathbf{m}$ in Eq. \ref{eq:cadherin_sindy} has a similar shape. Note that a commutator $\{E, m\}$ can also occur as part of the convective derivative. However, because the per-edge levels of myosin $m$ transform as a density (myosin is diluted if an edge is stretched), such a convective term would have the opposite sign as strain rate feedback, which is a form of ``anti-dilution''. \\

\subsubsection{Discrepancy of SINDy model with experiment due to different types of tissue strain}
While a strain-rate term does appear in the SINDy equations, its dorso-ventral modulation behaves unexpectedly. Based on the experimental results explained above, we were expecting high levels of strain-rate ventro-laterally, and low levels above. However, the E-cadherin-dependent strain-rate coefficient in Eq. \ref{eq:myosin_sindy} is $(1 + a_E\, c)$ with $a_E>0$, so that the opposite is true: in the SINDy model, the strain rate feedback is strongest dorsally.Furthermore, in the SINDy model, strain rate feedback does not significantly contribute to the establishment of the DV myosin gradient, which is opposite to the results of Ref. \cite{gustafsonPatternedMechanicalFeedback2022}.
Instead, our learned equations generate a DV myosin gradient through excitable tension dynamics which are controlled by the E-cadherin field.
To understand the discrepancy between these two pictures, we need to revisit how strain can appear in a tissue.

In the absence of cell division and death, a tissue can either deform by stretching individual cells, and hence junctions, or by cell rearrangement without changing cells shapes \cite{blanchardTissueTectonicsMorphogenetic2009}. 
Ref.~\cite{gustafsonPatternedMechanicalFeedback2022} established a feedback only in response to cell stretching.
Cell rearrangement will not induce a similar feedback, because during cell rearrangements individual junctions are not stretched. 

Immediately before \emph{Drosophila} GBE,  mesodermal cells along the ventral pole invaginate, forming the VF.  This causes cells in the ventral lateral domain of the germband to stretch, driving a temporary tissue flow towards the ventral pole.  Subsequently, flow is dominated by cell rearrangements \cite{brauns_epithelial_2023}.
Our coarse-grained flow field, which is computed from image data using PIV and further processed via smoothing and PCA, omits the cell-scale information needed to distinguish between these two modes of driving tissue flow --- cell stretching and cell rearrangement.

This lack of distinction between rearrangement and cell shape provides a possible explanation for the discrepancy between~\cite{gustafsonPatternedMechanicalFeedback2022} and our machine-learned strain rate term.
The $(1 + a_E\, c)$ E-cadherin dependence learned by SINDy reduces feedback in the lateral regions, where strain-rate during GBE primarily comes from rearrangement~\cite{brauns_epithelial_2023}.
Strain-rate feedback in the lateral region would kill the myosin field and stop flow, because there, strain rate $E$ and myosin orientation are opposite.
It strengthens feedback in the dorsal tissue, which deforms via both rearrangement and deformation~\cite{brauns_epithelial_2023}.
The role of our machine-learned feedback, which appears to prioritize deformation and ignore rearrangement, is to preserve myosin at the dorsal pole at late times.
We note that this also suppresses steep myosin gradients which would cause the tissue flow to continue to grow.
Such an effect is consistent with our earlier analysis which uses minimal hydrodynamic models to show that mechanical feedback can help stabilize flow patterns during GBE.

\subsection{Analysis of experimental data for Figures 4-6}

In this section, we describe the analysis of the experimental data shown in main text Figs. 4-6. We used slightly different image processing protocols than for the preceding Figures 1-3, where the main aim is to pre-process data for machine learning models.

\subsubsection{Analysis of \emph{Drosophila} E-cadherin and mysoin levels in DV and AP mutants}

Pullback images were created from 3d light sheet data using tissue cartography as described in Materials \& Methods. To time-align movies from DV mutants, we used the formation of the cephalic furrow and the invagination of the posterior midgut, all of which are preserved in DV patterning mutants. For AP mutants (Fig. 5), we used PIV to time align. Since the overall level of flow is very low, the uncertainty due to this manual alignment was not a serious concern. Data from fixed embryos (E-cadherin) was aligned by comparing the morphology, in particular the characteristic folds, with the live movies.

We pre-processed the myosin and E-cadherin images as follows. Myosin was analysed using the cytosolic normalization filter described above. E-cadherin has very low levels in the cytosol, so the cytosolic normalization cannot be used. Instead, we normalized the contrast of all images by setting the $5^\text{th}$ and $95^\text{th}$ percentiles to $0$ and $1$, respectively. Note that therefore, comparisons of absolute levels of E-cadherin are not possible, in contrast to myosin.
To measure myosin and E-cadherin gradients, we averaged the signal over the AP axis in a rectangular window that spanned the distance between the cephalic furrow and the beginning of the region that invaginates to form the posterior midgut. Tissue flow was analysed using PIV as described above.

\begin{figure}
    \centering
    \includegraphics[width=0.7\textwidth]{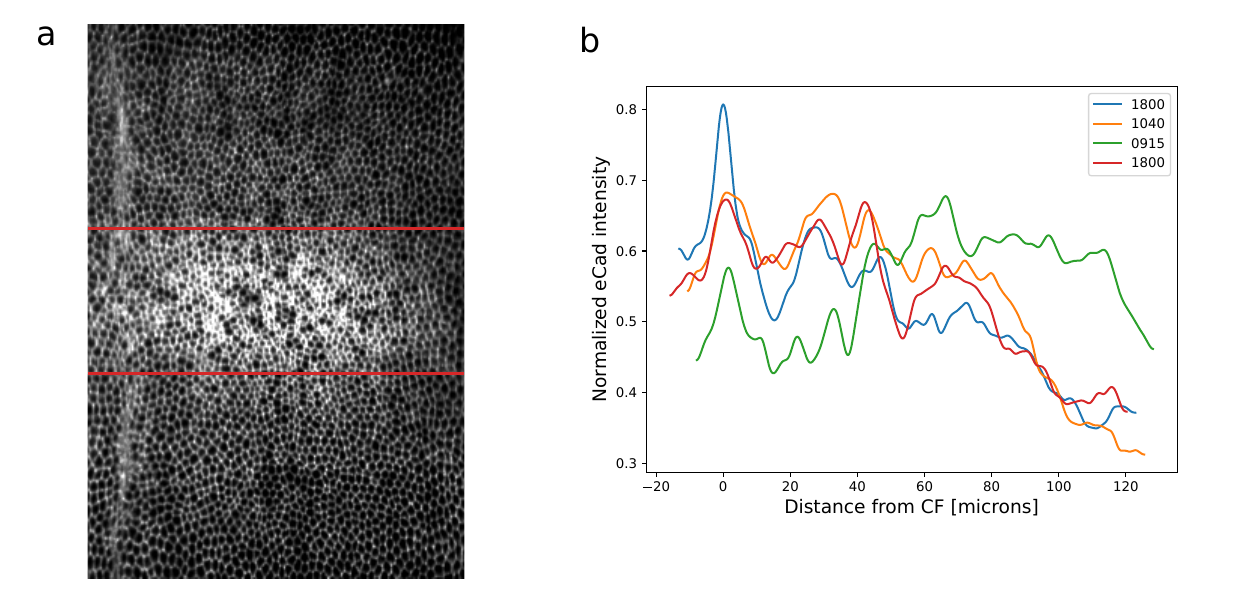}
    \caption{AP-variation of dorsal E-Cadherin gradients along single stripes.
    (a-a') E-Cadherin and region analyzed in (b). Timing is 5 minutes before VF initiation
    (b) E-Cadherin AP traces of $N=4$ embryos.
    }
    \label{fig:ECad-AP}
\end{figure}

\begin{figure}
    \centering
    \includegraphics[width=0.3\textwidth]{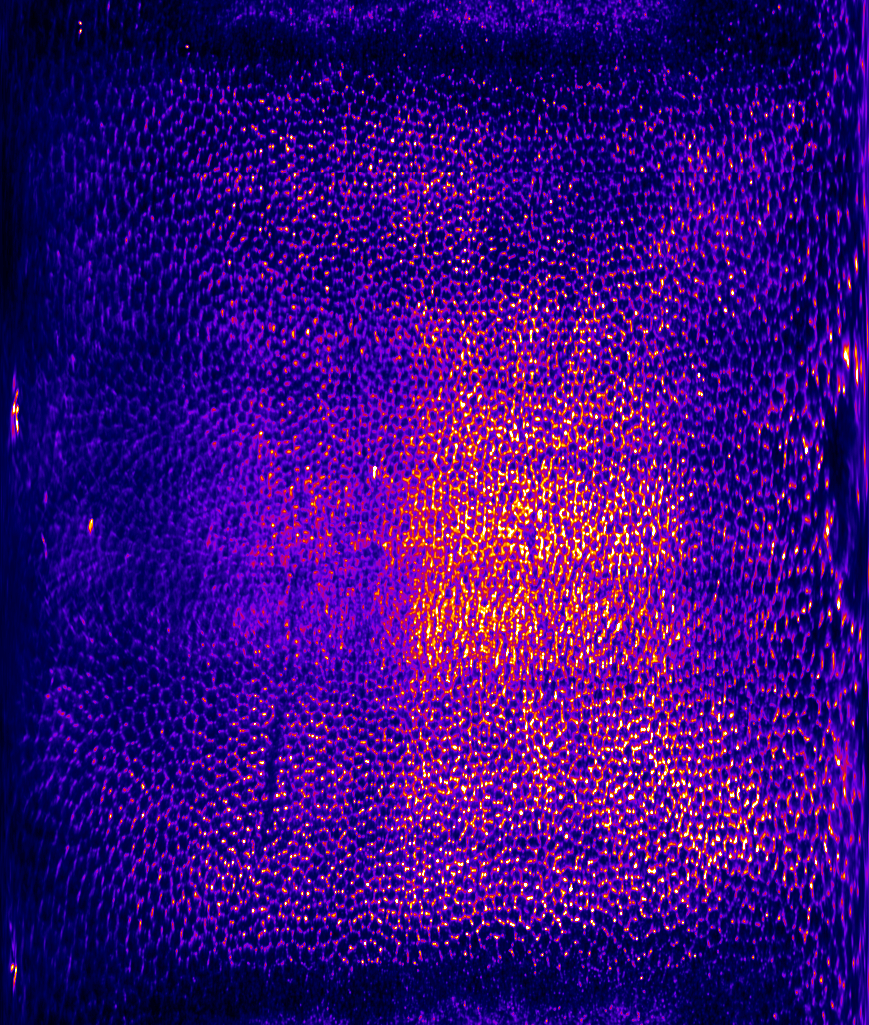}
    \caption{Bazooka/PAR3 levels are graded along the DV axis. False-color image of immunostaining of Bazooka at the onset of GBE.
    }
    \label{fig:bazooka}
\end{figure}

We also analyzed the AP-variation of myosin and E-cadherin. As examined in Ref.~\cite{lefebvre_geometric_2023}, there are no systematic AP stripes in the myosin pattern. The AP-variation of E-cadherin is quantified in Fig.~\ref{fig:ECad-AP}.

%Finally, 

\subsubsection{Analysis of \emph{Drosophila} AP genetic patterning}

\begin{figure}
    \centering
    \includegraphics[width=0.7\textwidth]{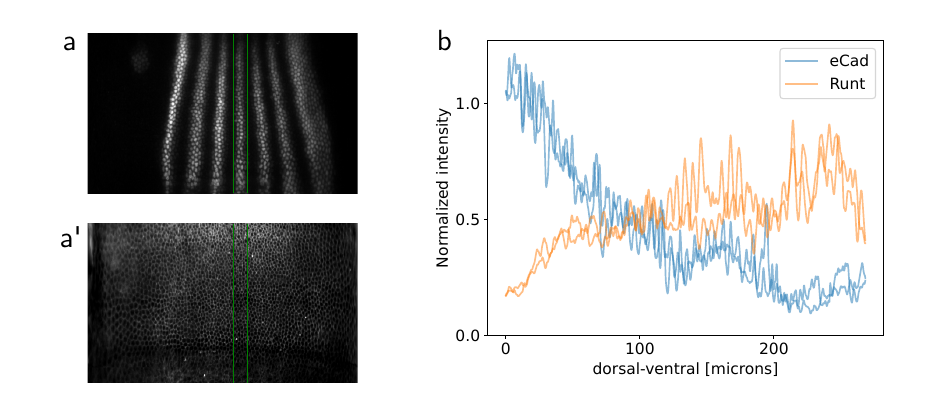}
    \caption{DV gradients along single stripes.
    (a-a') E-cadherin and Runt images analysed in (b). Traces correspond to Runt stripe marked in green. One lateral side is shown, but both were analyzed.
    Both images are 5 minutes before VF initiation
    (b) DV traces of Runt and E-cadherin image shown in (a-a').
    }
    \label{fig:single-stripe}
\end{figure}

To analyze the gradients of homeobox genes, we normalized the images by setting the $5^\text{th}$ and $95^\text{th}$ percentiles to $0$ and $1$, respectively. We then  we averaged the signal over the AP axis in a curved window that spanned the distance between the $2^\text{nd}$  and $7^\text{th}$ Runt stripes. We adjusted the signal by the cell density, which is higher on the dorsal side of the embryo. We applied the same procedure to the TLRs Tartan and Toll-6 shown in Fig. 5.  Note also that in the region we analyze -- between Runt stripes 1 and 7, excluding the anterior and posterior poles -- the diameter of the embryo varies very little~\cite{mitchellMorphodynamicAtlasDrosophila2022} and curvature is negligible.
To show that the gradients in PRGs and E-cadherin are not artifacts of the smoothing and AP-averaging procedure, we show below in SI Fig.~\ref{fig:single-stripe} the DV-profile for a single Runt stripe, without smoothing.
The AP-variation of E-cadherin is quantified in Fig.~\ref{fig:ECad-AP}.

\subsubsection{Analysis of neural tube organoids}

All images of neural tube organoids (NTOs) were taken on a Leica SP8 confocal microscope as $z$-stacks. For this system, time alignment as well as the morphogenetic process are controlled by the addition of BMP4. In all quantifications we exclude the ``dome'', i.e. the single-cell layer surrounding the outer lumen, making the cutoff where the cells do not adhere to the glass substrate anymore. We performed three different kinds of image analyses on these datasets. 
\begin{enumerate}
    \item For myosin and E-cadherin as shown in Fig. 6, we used surface extraction and tissue cartography to extract the apical surface of the NTO. For myosin, we used these surfaces to quantify the radial gradient. In the case of myosin, we found fluorescent granules in the NTO lumen, close to the apical surface. We removed these granules computationally where possible, and excluded regions where the granules were too close to the apical surface to be separated from it from our quantifications.
    \item For E-cadherin, we used a slightly different method to quantify the radial gradient which was more convenient for a large number of samples. We created radial slices, as shown in Fig. 6b, rotating the sample by $45$ degree increments to obtain multiple perspectives, and then extracted the apical surface within such a 2d slice. We then quantified the E-cadherin levels along the extracted surface. 
    \item For dataset showing nuclear-localized markers (Hes1, Zic2, Smad 1/5/9), we used the DAPI channel to create a nuclear mask, excluding all non-nuclear signal. We then created radial slices as before, and computed the average signal, normalized by nuclei area, along the $z$-axis (vertical axis in Fig. 6).
\end{enumerate}
For all three analyses, we standardized the images by setting the $5^\text{th}$ and $95^\text{th}$ percentiles to $0$ and $1$, respectively.

\begin{figure}
    \centering
    \includegraphics{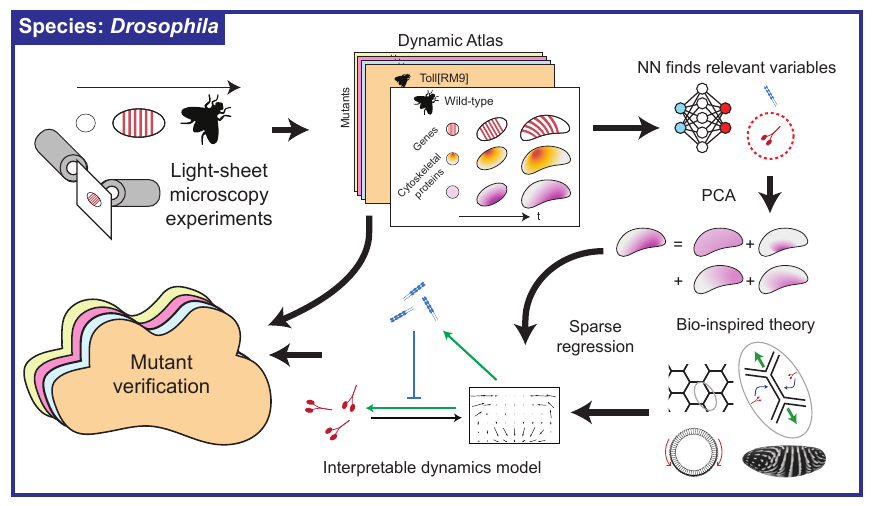}
    \caption{Workflow for data-driven analysis. NNs mined the morphodynamic atlas to determine the most relevant fluorescent-imaged proteins for predicting tissue flow. Using PCA and sparse regression informed by active matter theory, we developed an interpretable model for the coupled dynamics of two cytoskeletal proteins and tissue flow. This model produced qualitative predictions about the effect of the early-time E-cadherin pattern which we tested by analyzing mutant embryos.}
    \label{fig:ml_workflow}
\end{figure}

%\end{comment}

\end{document}